\UseRawInputEncoding
\documentclass[reqno,12pt,letterpaper]{amsart}
\usepackage{amssymb}
\usepackage{graphicx}
\usepackage{tensor}
\usepackage{mathrsfs}
\usepackage{slashed}
\usepackage{amsmath,amssymb,amsthm,graphicx,mathrsfs,url}

\usepackage[margin=1in, marginpar=0.6in]{geometry} 

\usepackage[dvipsnames,svgnames,table]{xcolor}
\usepackage[colorlinks = true,linkcolor = BrickRed,citecolor = Green, pdfencoding=auto, psdextra, bookmarks=true, bookmarksopen=true, pdfpagemode=UseOutlines]{hyperref}

\hypersetup{linktocpage}

\usepackage{enumitem}

\allowdisplaybreaks				

\usepackage{mathtools}
\mathtoolsset{showonlyrefs}

\definecolor{green}{rgb}{0,0.8,0} 

\newtheorem{theorem}{Theorem}[section]
\newtheorem{corollary}[theorem]{Corollary}
\newtheorem{lemma}[theorem]{Lemma}
\newtheorem{proposition}[theorem]{Proposition}
\theoremstyle{definition}
\newtheorem{definition}[theorem]{Definition}

\theoremstyle{remark}
\newtheorem{remark}[theorem]{Remark}
\numberwithin{equation}{section}
\newcommand{\relphantom}[1]{\mathrel{\phantom{#1}}}

\makeatletter
\newcommand{\nrm}{\@ifstar{\nrmb}{\nrmi}}
\newcommand{\nrmi}[1]{\Vert{#1}\Vert}
\newcommand{\nrmb}[1]{\left\Vert{#1}\right\Vert}
\newcommand{\abs}{\@ifstar{\absb}{\absi}}
\newcommand{\absi}[1]{\vert{#1}\vert}
\newcommand{\absb}[1]{\left\vert{#1}\right\vert}
\newcommand{\brk}{\@ifstar{\brkb}{\brki}}
\newcommand{\brki}[1]{\langle{#1}\rangle}
\newcommand{\brkb}[1]{\left\langle{#1}\right\rangle}
\newcommand{\set}{\@ifstar{\setb}{\seti}}
\newcommand{\seti}[1]{\{#1\}}
\newcommand{\setb}[1]{\left\{ #1\right\}}
\makeatother

\newcommand{\td}[1]{\widetilde{#1}}
\newcommand{\br}[1]{\overline{#1}}

\newcommand{\wh}[1]{\widehat{#1}}

\newcommand{\VERT}[1]{{\left\vert\kern-0.25ex\left\vert\kern-0.25ex\left\vert #1 
    \right\vert\kern-0.25ex\right\vert\kern-0.25ex\right\vert}}

\DeclareMathOperator{\dist}{dist}
\DeclareMathOperator{\sgn}{sgn}
\DeclareMathOperator{\supp}{supp}

\let\Re\relax
\DeclareMathOperator{\Re}{Re}
\let\Im\relax
\DeclareMathOperator{\Im}{Im}

\newcommand{\aeq}{\simeq}
\newcommand{\aleq}{\lesssim}

\newcommand{\lap}{\Delta}

\newcommand{\ud}{\mathrm{d}}

\newcommand{\0}{\emptyset}

\newcommand{\peq}{\relphantom{=}}			

\newcommand{\alp}{\alpha}
\newcommand{\bt}{\beta}
\newcommand{\gmm}{\gamma}

\newcommand{\dlt}{\delta}
\newcommand{\Dlt}{\Delta}
\newcommand{\eps}{\epsilon}

\newcommand{\kpp}{\kappa}
\newcommand{\lmb}{\lambda}
\newcommand{\Lmb}{\Lambda}
\newcommand{\vphi}{\varphi}

\newcommand{\tht}{\theta}

\newcommand{\vtht}{\vartheta}
\newcommand{\omg}{\omega}
\newcommand{\Omg}{\Omega}

\newcommand{\zt}{\zeta}

\newcommand{\bfr}{{\bf r}}

\newcommand{\bfT}{{\bf T}}

\newcommand{\bbC}{\mathbb C}

\newcommand{\bbR}{\mathbb R}
\newcommand{\bbS}{\mathbb S}

\newcommand{\calC}{\mathcal C}
\newcommand{\calD}{\mathcal D}

\newcommand{\calH}{\mathcal H}

\newcommand{\calM}{\mathcal M}

\newcommand{\calU}{\mathcal U}
\newcommand{\calV}{\mathcal V}

\newcommand{\frkn}{\mathfrak n}

\newcommand{\frkr}{\mathfrak r}

\DeclareMathAlphabet{\mathbbold}{U}{bbold}{m}{n}	
\newcommand{\R}{{\mathbb R}}
\newcommand{\p}{\partial}
\newcommand{\scrS}{\mathscr S}

\newcommand{\bop}{{\mathrm{b}}}

\newcommand{\Arg}{{\mathrm{Arg}}}
\newcommand{\QNM}{{\mathrm{QNM}}}
\newcommand{\scop}{{\mathrm{sc}}}

\newcommand{\rst}{|}			

\usepackage{titletoc}
\newcommand{\nocontentsline}[3]{}
\let\origcontentsline\addcontentsline
\newcommand\stoptoc{\let\addcontentsline\nocontentsline}
\newcommand\resumetoc{\let\addcontentsline\origcontentsline}

\begin{document}

\title[QNM for multiple black holes]{Long lived QNM for multiple black holes}

\author{Ning Tang}%
\address{Department of Mathematics, UC Berkeley, Berkeley, CA 94720, USA}%
\email{ning\_tang@math.berkeley.edu}%

\author{Zhongkai Tao}
\address{Institut des Hautes \'{E}tudes Scientifiques, 91440 Bures-sur-Yvette, France}
\email{ztao@ihes.fr}


\begin{abstract}
\emergencystretch=2em
    We provide a rigorous definition of quasinormal modes (QNM) for the Majumdar--Papapetrou spacetimes with multiple extremal black holes. For certain configurations,
    the trapped set of the null geodesic flow contains stable (elliptic) orbits. We analyze this phenomenon in the case of two black holes of equal mass and show that it produces QNM converging exponentially fast to the real axis. QNM for black holes have been extensively studied for {\em single} black holes and this appears to be the first {\em mathematical} result for the case of multiple black holes.
\end{abstract}
\maketitle
\setcounter{tocdepth}{1}
\titlecontents{section}
[0em]
{\vspace{0.4em}}
{\contentslabel{2em}}
{\bfseries}
{\bfseries\titlerule*[0.5pc]{$\cdot$}\contentspage}
[\vspace{0.2em}]
\titlecontents{subsection}
              [1.5em]
              {}
              {\contentslabel{2em}}%
              {}
              {\titlerule*[0.5pc]{$\cdot$}\contentspage}%
\titlecontents{subsubsection}
              [2.5em]
              {}
              {\contentslabel{2em}}%
              {}
              {\titlerule*[0.5pc]{$\cdot$}\contentspage}%


\section{Introduction}

The Majumdar--Papapetrou (MP) spacetime $(\calM^{1+3}, g_{MP})
$ is a solution to the Einstein--Maxwell equations modeling multiple black holes located at $ \mathbf r_j \in \mathbb R^3 $, $ j = 1, \cdots , N $. It is simply given by 
 \begin{equation}\label{eq:gMP}
    g_{MP} = -U^{-2}(x)\, \ud t^2 + U^2(x)\, \ud x^2, 
    \qquad 
     \Delta U ( x )= 0 , \ \ \ U ( x ) > 0 , \  \ \  x \in \calM := \R^3 \setminus \set{\bfr_j}_{j = 1}^N. 
\end{equation}
The specific harmonic function $ U $ is chosen so that $ N =1 $  corresponds to the extremal Reissner--Nordstr\"om (eRN) black holes \cite{Maj47, Pap47}:
\begin{equation}\label{eq:U}
    U(x) = 1 + \sum_{j = 1}^N \frac{M_j}{|x - \bfr_j|},\quad M_j>0.
\end{equation}
\begin{figure}[htbp]
    \centering
        \centering
        \includegraphics[width=16cm]{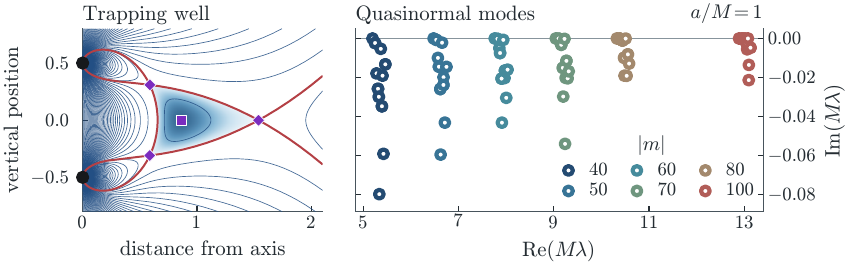} 
        \caption{QNM for two black holes of equal mass. Left: Level sets of the effective potential in cylindrical coordinates -- the well in the middle produces an elliptic orbit. Right: Numerically computed QNM for different values of the angular momentum $ m$ -- see Figures~\ref{fig:mp-kim-pasciak-resonances},\ref{fig:mp-height-contours} for more detailed presentations and 
        \href{https://math.berkeley.edu/~ztao/mp_movie_1.mp4}{link} for an animation showing a larger range of black hole separations.
        \label{fig:mp-frontpage}}
\end{figure}

The spectral family associated to $g_{MP}$ is defined as: 
\begin{equation}\label{eq:Plmb}
    P(\lmb) := -e^{i\lmb t}\Box_{g_{MP}} e^{-i\lmb t} = -U^{-2} \lap - \lmb^2 U^2 : C^\infty(\calM) \to C^\infty(\calM). 
\end{equation}
We prove the meromorphic continuation of 
\[ P ( \lambda )^{-1} : C_{\rm{c}}^\infty ( \mathcal M ) \to 
C^\infty ( \mathcal M ) , \]
so that QNM are characterized as finite-rank poles of this meromorphic family of operators. (In our informal convention, accepted in math and physics, we refer to both the frequencies and the corresponding states as modes.) 
\begin{theorem}\label{thm:meromorphic-ext}
 The inverse of the spectral family \eqref{eq:Plmb},
    \begin{equation}
        R(\lmb) := P(\lmb)^{-1}:L^2_{\mathrm{comp}}(\calM) \to L^2_{\mathrm{loc}}(\calM), \qquad  \Im \lmb > 0,
    \end{equation}
    has a meromorphic continuation to a conic neighborhood of the real axis in the lower half-plane.
    More precisely for any $\chi \in C^\infty_{\rm{c}}(\calM)$,  \[
        \chi P(\lmb)^{-1} \chi: L^2(\calM) \to L^2(\calM)
    \]
    extends from $\set{\Im \lmb > 0}$ to 
    \[
        \bbC \setminus i(-\infty, 0] = \left\{\lmb \in \bbC \setminus \set{0}: \arg(\lmb) \in \left(- \tfrac{1}{2}\pi,  \tfrac{3}{2} \pi \right)\right\}
    \]
    as a meromorphic family of bounded operators with poles of finite rank; the poles form a discrete subset independent of $\chi \not \equiv 0$. 
\end{theorem}
We denote the discrete set of poles of $ \chi P ( \lambda)^{-1} \chi $,  $\QNM(\bfr_1, M_1, \ldots, \bfr_N, M_N)$.
The operator $P(\lmb)$ becomes elliptic in the scattering pseudodifferential calculus after complex scaling. Using that, the proof of Theorem~\ref{thm:meromorphic-ext} is carried out in Section~\ref{sec:meromorphic-cont}.
For a weaker statement of the theorem, a simpler proof is outlined in Remark~\ref{rmk:tht0-simpler-proof}.

We now specialize to the case of equal-mass binary black holes: $M_1 = M_2$, $ | \mathbf r_1 - \mathbf r_2 | = a $. 
The effective parameter is then $a/M$. 
Without loss of generality, we set $\bfr_1 = (0, 0, {a}/{2})$ and $\bfr_2 = (0, 0, -{a}/{2})$. We show that 
in a certain range (see \eqref{eq:rangeaM} below) there exist
elliptic trapped null geodesics. That implies that QNM can converge exponentially to the real axis. We stress that this is a result about genuine QNM (defined using Theorem \ref{thm:meromorphic-ext}) not merely approximate solutions to the spectral equation (see Proposition~\ref{prop:quasimodes}).
\begin{theorem}\label{thm:exp-conv-QNM}
    Set $N = 2$, $\bfr_1 = (0, 0, {a}/{2})$ and $\bfr_2 = (0, 0, -{a}/{2})$ and $M := M_1 = M_2$. Assume that
    \begin{equation}
        \label{eq:rangeaM}
   \frac{4}{\sqrt{27}}  < \frac a M  < \frac{4\sqrt{2}}{\sqrt{27}} . 
   \end{equation}
    Then there exist $E>0$ and $c>0$ such that for every $\delta>0$, there exists a sequence $\set{z_m}_m \subseteq \QNM(\bfr_1, M, \bfr_2, M) $ for $m\geq m_0$ such that \[
    (\sqrt{E}-\delta)m < \Re z_m < (\sqrt{E}+\delta)m, \qquad 0 < -\Im z_m < e^{-cm}.
    \]
\end{theorem}

The range \eqref{eq:rangeaM} was discovered by Shipley--Dolan \cite[Section III.A, Appendix B]{ShipleyDolan16} through analysis in the equatorial plane (see that paper for references to earlier work in physics). A finer analysis of bounded orbits was provided by Nakashi--Igata~\cite[Section III, Appendix]{NI19}. Here we obtain \eqref{eq:rangeaM} through a detailed analysis of the effective potential -- see Proposition~\ref{prop:equal-mass-V-localmin}. For spacetimes with more than two black holes, the conclusion of Theorem~\ref{thm:exp-conv-QNM} is expected to hold if the property of the effective potential in Proposition~\ref{prop:equal-mass-V-localmin} remains valid.

The decay rate of imaginary parts is optimal as shown by results of
Burq \cite{Burq02} and Cardoso--Vodev \cite{CardosoVodev02} (see also \cite[Theorem 6.26]{DyZw19} for a textbook presentation in a simplified setting).

\begin{figure}[htbp]
    \centering
      \renewcommand{\thempfootnote}{\arabic{mpfootnote}}
        \centering
        \includegraphics
        [width=17cm]
        {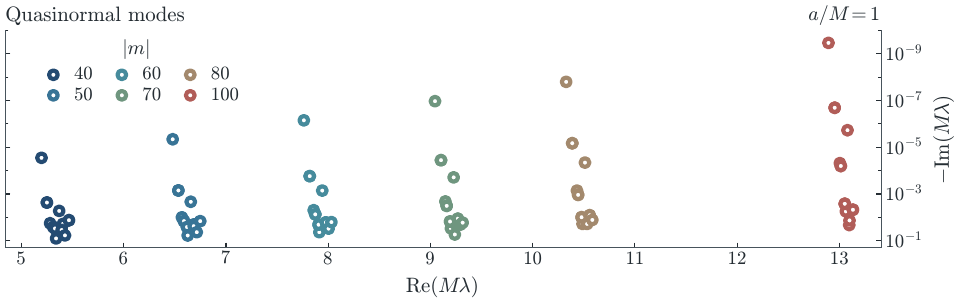}
        \caption{Numerically
    computed QNM in the setting of Theorem~\ref{thm:exp-conv-QNM} with $a/M=1$. We show resonances from Figure~\ref{fig:mp-frontpage} with imaginary parts on a logarithmic scale: the exponential convergence to the real axis is strikingly visible. The numerical schemes used here are briefly described in \S \ref{s:nume}.
    \label{fig:mp-kim-pasciak-resonances}}
           \end{figure}

The proof of Theorem~\ref{thm:exp-conv-QNM} is based on the  construction of approximate solution to $ P ( \lambda ) u = 0$. Following the classical work of Ralston \cite{Ralston76}, they can be used to show slow decay of waves and hence might be of independent interest.
\begin{proposition}\label{prop:quasimodes}
    The following holds under the assumptions of Theorem~\ref{thm:exp-conv-QNM}. There exist $E > 0$, $c>0$, and a bounded region $\Omg \Subset \R_{\rho > 0} \times \R_z$ such that for every $\dlt > 0$, there exists $m_0$ such that there exists a sequence $\lmb_m \in \R$ and $v_m \in L^2(\calM)$, $m \geq m_0$ such that 
    \begin{enumerate}[leftmargin = *]
        \item $(\sqrt{E} - \dlt)m < \lmb_m < (\sqrt{E} + \dlt)m$, 
        \item $\supp v_m \subseteq \Omg \times [0, 2\pi)_\vtht$, and 
        \item the functions $v_m$ are normalized exponentially accurate quasimodes in the sense that \[
            \|v_m\|_{L^2(\calM)} = 1, \qquad \|P(\lmb_m) v_m\|_{H^k(\calM)} \aleq_k e^{-cm}.
        \]
    \end{enumerate}
\end{proposition}

\subsection{Related works}\label{subsec:related-works}
MP spacetime is a special class of static solutions to the Einstein--Maxwell equations with $N$ extremal black holes, each with vanishing surface gravity (see Aretakis \cite{Aretakis15}). Its geometry was first studied in Hartle--Hawking \cite{HartleHawking72}. Chru\'sciel--Tod \cite{ChruscielTod07} later proved that the only asymptotically flat regular static black hole exteriors in $4$ dimensional Einstein--Maxwell theory are the (subextremal/extremal) RN or the standard MP class, which extends an earlier result of Chru\'sciel--Nadirashvili \cite{ChruscielNadirashvili95} within the MP class. A higher dimensional analog of MP spacetime was first found by Myers \cite{Myers87}, whose classification results were established in Lucietti \cite{Lucietti21}. For dynamical aspects of extremal black holes, we refer to the review articles by Hintz--Holzegel \cite{HintzHolzegel} and Dafermos \cite{Dafermos25}, and to the references therein.

Mathematical study of QNM was initiated by Bachelot--Motet-Bachelot \cite{BMB93}. It remains a topic of interest in physics and mathematics -- see for instance Assumpcao et al. 
\cite{ACIRZ18} and Hitrik--Zworski \cite{HZ25}, respectively. A survey by Berti et al. \cite{Berti_2026} can be consulted for a broad modern physics perspective. Below we will review results which are directly relevant to our results and proofs.

\subsubsection{QNM on black hole spacetimes}
Mathematical literature has been exclusively devoted to different kinds of {\em single} black holes. The MP model for one black hole 
is the eRN black hole. The QNM in that case were studied by 
Gajic--Warnick \cite{GajicWarnick21} who characterized quasinormal modes 
as eigenvalues of infinitesimal generators (with an appropriate restriction) of the solution operator to the wave equation. Moreover, the relation to scattering resonances was also established in a suitable subset of the complex plane; see \cite[Theorem 4.2]{GajicWarnick21}. 

Convergence of QNM to the real axis (as in Theorem \ref{thm:exp-conv-QNM}) has been proved only in the asymptotically Anti-de Sitter setting -- see Gannot \cite{Gannot14,Gannot17} for Kerr--AdS and Kehle \cite{Kehle21} for RN--AdS. An analog of Proposition \ref{prop:quasimodes} and of its consequences for the decay of waves was investigated in that setting by Holzegel--Smulevici \cite{HolzegelSmulevici13,HolzegelSmulevici14}. See also Keir \cite{Keir16}, Benomio \cite{Benomio}, Kehle \cite{Kehle22} and Kehle--Moschidis \cite{KehleMoschidis26}.

Since the effective operator in \eqref{eq:conjugation-h} is non-negative, there are no mode stability issues in our setting. We have not, however, investigated the behavior at $ 0 $ yet. For recent
advances about the behavior at $ 0 $, see Stucker \cite{stucker2024,Stucker26-Schw} and references given there.

\subsubsection{General theory of scattering resonances}\label{subsubsec:distributionQNM}

The relation between the distribution of scattering resonances and
classical dynamics, and trapping in particular, was emphasized early on by Lax--Phillips  \cite{LaxPhillips89} -- see \cite[\S\S 6.1, 6.6, 7.1, 7.6]{DyZw19} for a modern perspective and references. The relevant
results about passing from approximate solutions to resonances come from 
Tang--Zworski \cite{TangZworski} and its refinement by Stefanov \cite{Stefanov99}. 

\subsubsection{Quasimode construction along elliptic trajectories}
Construction of approximate solutions associated to elliptic trajectories has a long tradition going back to Babich, Lazutkin and Ralston. More recent references include the work of 
Cardoso--Popov \cite{CardosoPopov02} (construction of quasimodes of
subexponential accuracy $ \exp ( - c h^{-1+\delta} ) $) and  Christianson \cite{Christianson11} (a flexible quantum-monodromy approach to obtain $O(h^\infty)$ quasimodes). It seems that (in the setting of analytic metrics) general constructions of exponentially accurate quasimodes is not known. Here by quasimodes we mean approximate eigenvalues as in Proposition \ref{prop:quasimodes}.

When a symmetry reduction results in a Schr\"odinger operator with a well, exponentially accurate quasimodes can be constructed and in the context of GR that was done in the works  \cite{HolzegelSmulevici14, Keir16, Benomio} cited above. The key component are Agmon estimates -- see \cite[\S 6.2]{TaoZworski} for a friendly introduction.

\subsection{Numerical Experiments}
\label{s:nume}
The experiments have been performed by ChatGPT and were not independently verified by the authors.
ChatGPT used Python to approximate QNM for $|m| = 40,50,60,70,80,100$ from 
\eqref{eq:Plmb}. It used degree-$8$ conforming finite elements and a frequency-independent radial perfectly matched layer adapted from \cite{KimPasciak2009}. The $90$ modes shown passed frequency and eigenfunction-overlap checks under discretization and layer refinements. A separate sanity check in the $N=1$ case yielded an approximate lattice consistent with \cite[IV]{FerrariMashhoon84}.

\stoptoc
\subsection*{Acknowledgment}
The authors would like to thank Peter Hintz for bringing \cite{ShipleyDolan16} to their attention, and Sung-Jin Oh and Maciej Zworski for their interest in this work and stimulating discussions. The authors would also like to thank Maciej Zworski for encouraging us to prepare the figures in this paper, for helpful comments on a preliminary version of the manuscript, and for his help in improving the introduction. N.T. also thanks Ryan Unger for helpful discussions on extremal black holes. N.T. acknowledges support from Sung-Jin Oh's  National Science Foundation Grant under DMS-2452760. Z.T.~was partially supported by a grant from the National Research Agency (ANR) under the France 2030 program, reference ANR-25-CFFS-0012.

\subsection*{AI Statement}
\begin{figure}[htbp]
    \centering
    \begin{minipage}{\linewidth}
    \renewcommand{\thempfootnote}{\arabic{mpfootnote}}
        \centering
        \includegraphics[width=0.9\linewidth]{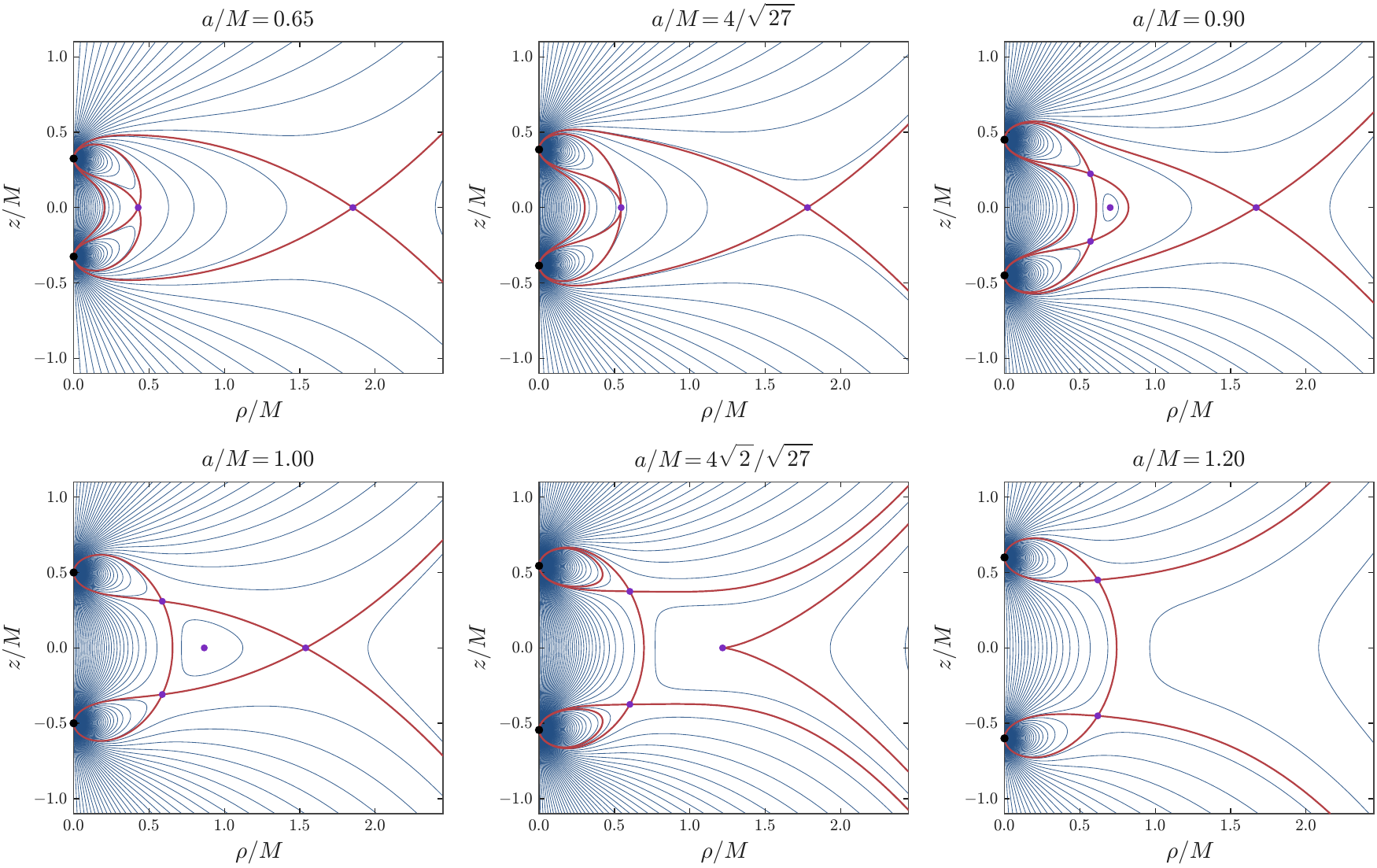}
        \input{MP_height_caption.tex}
        \label{fig:mp-height-contours}
    \end{minipage}
\end{figure}

An initial proof of Proposition~\ref{prop:equal-mass-V-localmin} was independently generated by ChatGPT during discussions initiated by the authors, which was subsequently checked and refined by the authors. Figures~\ref{fig:mp-frontpage}, ~\ref{fig:mp-kim-pasciak-resonances},~\ref{fig:mp-height-contours},~the animation~\href{https://math.berkeley.edu/~ztao/mp_movie_1.mp4}{link} and the account of their generation procedure were generated by ChatGPT. The property \eqref{eq:conjugation-h} was pointed to us by ChatGPT.
ChatGPT was also used for exploratory mathematical discussions, locating references, understanding arguments in the literature, proofreading and language editing, and drafting this statement. AI use was limited to these activities; apart from the assistance explicitly mentioned above, no content of this paper was generated by AI. 

\resumetoc

\section{Preliminaries}


\subsection{Coordinate systems}

\subsubsection{Standard Euclidean coordinates}
We introduce globally defined rectangular coordinates $(t = x^0, x^1, x^2, x^3)$ using the diffeomorphism $\calM^{1+3} \to \R \times (\R^3 \setminus \set{\bfr_j}_{j = 1}^N)$. We will use $\p_\mu = \p_{x^\mu}$ to denote the coordinate vector field in this coordinate system.
With a slight abuse of notation, we also use $t = x^0, x = (x^1, x^2, x^3)$ and $x_j = x^j$.
With respect to this coordinate system, we define the smooth vector field \[
    \bfT = \p_t,
\]
which is a Killing vector field. Its level hypersurfaces are given by \[
    S_{t_0} := \set{p \in \calM : t(p) = t_0}.
\]
Using the global rectangular coordinates, we define a global function \[
    r = \sqrt{\sum_{a = 1}^3 (x^a)^2}.
\]

\subsubsection{Local coordinates near event horizons}\label{subsubsec-reg-coord}
Near each event horizon $\calH^+_j$ ($j = 1, 2, \ldots, N$), we introduce the radial coordinate
\[
    r_{\bop_j} := |x - \bfr_j|, \qquad j = 1, 2, \ldots, N.
\]
For notational convenience, we shall subsequently suppress the index $j$ -- the notation $(r_\bop, \calH^+, \bfr, M)$ will then be understood as referring to each corresponding pair $(r_{\bop_j}, \calH^+_j, \bfr_j, M_j)$.

The local spherical coordinates $(t, r_{\bop_j}, \omg)$ around $\calH^+$ are valid for $0 < r_{\bop_j} = |x - \bfr_j| < \min_{J \neq j} |\bfr_j - \bfr_J|$: \[
    x - \bfr =  r_\bop \omg, \qquad \omg = (\sin\vphi\cos\tht, \sin\vphi\sin\tht, \cos\vphi) \in \bbS^2.
\]

\subsubsection{Cylindrical coordinates}
When $N = 2$, $\bfr_1 = (0, 0, \tfrac{a}{2})$ and $\bfr_2 = (0, 0, -\tfrac{a}{2})$, we define the following functions \[
    \rho = \sqrt{(x^1)^2 + (x^2)^2}, \qquad (\cos \vtht, \sin\vtht) = (\tfrac{x^1}{\rho}, \tfrac{x^2}{\rho}) \in \bbS^1
\]
such that 
\begin{equation*}
    U(x) = 1+\frac{M_1}{\sqrt{\rho^2+(z-a/2)^2}}+\frac{M_2}{\sqrt{\rho^2+(z+a/2)^2}}.
\end{equation*}
In this coordinate system, the metric is of the form \[
    g = -U^{-2}\, \ud t^2 + U^2(x) \, (\ud \rho^2 + \rho^2 \ud \vtht^2 + \ud z^2).
\]

We refer to the coordinates $(t, \rho, z = x^3, \vtht)$ as the cylindrical coordinates on $\calM^{1+3}$.

\subsection{Resolvent}

Note that $-U^{-4}\lap: C^\infty_{\rm{c}}(\calM) \subseteq L^2(\calM; U^4\, \ud x) \to L^2(\calM; U^4\, \ud x)$ is dense, symmetric and nonnegative and hence admits a self-adjoint extension. 

Thanks to $\lap U = 0$, a direct computation (see \cite[Corollary 4.3]{MasonNicolasCST}) leads to \[
    \Box_{g_{MP}} = -U^2 \p_t^2 + U^{-2}\lap = U^3 \Box_{U^2(x)g_{MP}}U^{-1}.
\]
Thus, \begin{equation}\label{eq:conjugation-h}
    -U^{-2}\lap = -U^3\Delta_h U^{-1}, \qquad h=U^4 \ud x^2.
\end{equation}
Since $(\calM, U^4 \ud x^2)$ is complete, $-\Delta_h$ is essentially self-adjoint by \cite[\S 1.3 Theorem 3.1]{Shubin92}. Hence, $-U^{-4}\lap: C^\infty_{\rm{c}}(\calM) \subseteq L^2(\calM; U^4\, \ud x) \to L^2(\calM; U^4\, \ud x)$  is also essentially self-adjoint. Thus, by the spectral theorem and its nonnegativity, \begin{equation}\label{eq:U4-normalized-resolvent}
    (-U^{-4} \lap - \lmb^2)^{-1}: L^2(\calM; U^4\, \ud x) \to L^2(\calM; U^4\, \ud x)
\end{equation}
is well-defined and holomorphic for $\Im \lmb > 0$. Therefore, \begin{equation}\label{eq:Rlmb-Lmb0}
    R(\lmb) := P(\lmb)^{-1} 
    : L^2(\calM;  \ud x) \to L^2(\calM; U^4\, \ud x) \hbox{ is holomorphic for $\Im \lmb > 0$},
\end{equation}
where $P(\lmb)$ is given by \eqref{eq:Plmb}. 
In Section~\ref{sec:meromorphic-cont}, we work with $\td P(\lmb) = -\lap - \lmb^2 U^4(x)$ and establish the meromorphic continuation of $R(\lmb)$.

\subsection{Complex scaling}
The original mathematical treatment of complex scaling was due to \cite{AguilarCombes71}. We refer to, for example, \cite[Section 3]{SjZw91}, \cite[Section 4.5]{DyZw19}, for reviews on complex scaling. Now we recall some basic notions to prepare us for applying complex scaling in Section~\ref{sec:meromorphic-cont}.

\begin{definition}[{\cite[Definition 4.28]{DyZw19}}]
    An $n$-dimensional smooth submanifold $M$ of $\mathbb{C}^n \simeq \mathbb{R}^{2 n}$ is called a (maximally) totally real submanifold if for any $p \in M$, $T_p M \cap i T_p M=\{0\}$, where we identify the tangent space $T_p M$ with a subspace of $\mathbb{R}^{2 n} \simeq \mathbb{C}^n$. 
    
    If $u \in C^{\infty}(M)$ then $\td u \in C^{\infty}\left(\mathbb{C}^n\right)$ is called an almost analytic extension of $u$ if $\td{u}|_{M}=u$ and \[
        \bar{\partial}_{z_j} \tilde{u}(z)=\mathcal{O}\left(d(z, M)^{\infty}\right), \quad z \in \mathbb{C}^n .
    \]
\end{definition}
Given an immersion $F : \Omg \subseteq \R^n \to \bbC^n$ (i.e., $\ud F(p): \R^n \to \R^{2n}$ is injective), the criterion for $F(\Omg)$ being totally real is characterized by $(\ud F(p))(\R^n) \cap i(\ud F(p))(\R^n) = \set{0}$, where $T_{F(p)} F(\Omg) = \ud F(p)(\R^n)$.
Since $\ud F(p): \R^n \to \R^{2n} \aeq \bbC^n$ is a real linear injective map, it can be extended by linearity into a $\bbC$-linear map $\ud F(p): \bbC^n \to \bbC^n$. Then the condition of $F(\Omg)$ being totally real is \begin{equation}\label{eq:totally-real-udF}
    (\ud F(p))(\R^n) \cap (\ud F(p))(i\R^n) = \set{0} \iff \ud F(p) : \bbC^n \to \bbC^n \hbox{ is injective}.
\end{equation}

\begin{lemma}[{\cite[Lemma 4.30]{DyZw19}}]\label{lem:unique-ext-op-on-TRS}
    Suppose that $M$ is a totally real submanifold of $\bbC^n$. Then every $u \in C^{\infty}(M)$ has an almost analytic extension to $\mathbb{C}^n$ in the sense of above.

    If $\widetilde{P}=\sum_{|\alpha| \leq m} a_\alpha(z) \partial_z^\alpha$ is a holomorphic differential operator near $M$ (i.e., $a_\alpha$'s are holomorphic in $\mathbb{C}^n$ near $M$) then $\widetilde{P}$ defines a unique differential operator $P_M$ whose action on $C^{\infty}(M)$ is given by
    \[
        P_M u = \left.\widetilde{P}(\tilde{u})\right|_M .
    \]
\end{lemma}

In our application in Section~\ref{sec:meromorphic-cont}, our totally real submanifold is realized as the image of an immersion and hence the operator can be obtained explicitly by coordinate transformations; see \eqref{eq:tdP-bt-lmb}. In addition, we require the following deformation result.

\begin{lemma}[{\cite[Lemma 3.1]{SjZw91}}]\label{lem:deformation}
    Let $\Omg \subseteq \R^n$ be open, and let $f: [0,1] \times \Omg \ni(s, y) \mapsto f(s, y) =: f_s(y) \in \bbC^n$ be a smooth proper map such that $\det\left(\partial_y f(s, y)\right) \neq 0$ for all $(s, y)$ and such that $f_s$ is injective. Assume that $f_s = f_0$ on $\Omg \setminus K$ where $K$ is some compact subset of $\Omg$. Let $P\left(x, D_x\right)$ be a differential operator with holomorphic coefficients defined in a neighborhood of $f([0,1] \times \Omg)$ such that $P_{\Omg_s}$ is elliptic for $0 \leq s \leq 1$, where $\Omg_s := f_s(\Omg)$. If $u_0 \in \calD'\left(\Omg_0\right)$ and $P_{\Omg_0} u_0$ extends to a holomorphic function in a neighborhood of $f([0,1] \times \Omg)$, then the same thing holds for $u_0$.
\end{lemma}

\subsection{Scattering calculus}\label{subsec:sc}
In this subsection, we recall some basic properties for scattering calculus. We mainly follow \cite[Section 5]{Vasynotes} and \cite[Section 2]{UhlmannVasy16}.

We assume $M$ is an $n$-dimensional compact manifold with boundary. We define a subspace of vector fields $\calV(M)$: \[
    \calV_\bop(M) := \set{V \in \calV(M) : Vf|_{\p M} = 0 \hbox{ for } f|_{\p M} = 0}.
\]
Let $\rho$ be a boundary defining function such that $\rho \geq 0$ with equality taken exactly at $\p M$, and $\ud \rho \neq 0$ on $\partial M$. This allows us to introduce the scattering vector fields $\calV_\scop(M) := \rho\calV_\bop(M)$.
We begin with the scattering calculus on $\R^n$, following \cite{UhlmannVasy16}; see also \cite[Section 5.3.2]{Vasynotes}. We emphasize that our symbol class $S^{m,\ell}_\scop$ corresponds to $S^{m,\ell,0}_{\dlt=0,\dlt'=0}$ in \cite{Vasynotes}, rather than the larger class $S^{m,\ell,0}_{\infty,\dlt=0}$.

\begin{definition}
    We define the scattering symbols \begin{equation}\label{eq:defn-sc-symbol}
        S^{m, \ell}_\scop := \set{a \in C^\infty(\R^n_z \times \R^n_\zt) : |\p_z^\alp \p_\zt^\bt a| \leq C_{\alp\bt} \brk{\zt}^{m - |\bt|}\brk{z}^{\ell - |\alp|} }.
    \end{equation}
    One then defines $\Psi^{m, \ell}_\scop(\R^n)$ to consist of left quantizations of such symbols.
\end{definition} 
For such symbols, the principal symbols live in $S^{m,\ell}_\scop/S^{m-1, \ell-1}_\scop$. We say that $A \in \Psi^{m, \ell}_\scop(\R^n)$ is elliptic if there exists $b \in S^{-m, -\ell}_\scop$ such that $b\sigma(A) - 1 \in S^{-1, -1}_\scop$. The standard parametrix construction will give an operator $B \in \Psi^{-m, -\ell}_\scop$ such that $AB - I \in \Psi^{-\infty, -\infty}_\scop$, where the $\Psi^{-\infty, -\infty}_\scop$ is the residue class; see \cite[Section 5.3.7]{Vasynotes}. Note that an equivalent definition of ellipticity is that there exists $c, R > 0$ such that the symbol $a$ satisfies \begin{equation}\label{eq:sc-ell-defn}
    |a| \geq c \brk{z}^{\ell} \brk{\zt}^m, \qquad |z| + |\zt| > R.
\end{equation}

We introduce the weighted Sobolev spaces $H^{s, r}(\R^n) := \brk{z}^{-r}H^s(\R^n)$. Given $A \in \Psi^{m, \ell}_\scop(\R^n)$, it satisfies the mapping property $A: H^{s, r}(\R^n) \to H^{s - m, r - \ell}(\R^n)$.

Next, we will associate the discussion above on $\R^n$ with the geometric formulation on the radial compactification $\br{\R^n}$. It is straightforward to check that $\dot \calC^\infty(\br{\R^n}) = \scrS(\R^n)$, where $\dot \calC^\infty(\br{\R^n})$ is defined as smooth functions on $\br{\R^n}$ vanishing with all derivatives at $\p\br{\R^n}$. Moreover, we obtain a compactification $\br{\R^n_z} \times \br{\R^n_\zt}$ of the product space and we use $\rho_\p$ and $\rho_\infty$ for the global boundary defining function for the two boundaries.
It is again straightforward to check that the estimate in \eqref{eq:defn-sc-symbol} is equivalent to \[
    \rho_\infty^m \rho_\p^\ell V_1 V_2 \cdots V_k a \in L^\infty, \qquad V_j \in \calV_\bop(\br{\R^n_z} \times \br{\R^n_\zt}).
\]
The principal symbols are defined as the equivalence class in $S^{m, \ell}_\scop/ S^{m-1, \ell-1}_\scop$. We write $\Psi^{m, \ell}_\scop(\br{\R^n}) = \Psi^{m, \ell}_\scop(\R^n)$. 
In addition, we define $L^2_\scop(\br{\R^n})$ using the density given by a density $\br{\R^n}$-locally equivalent to the standard $L^2$ density on $\R^n$ in the sense that $x^{-n-1}|\ud x \ud y_1 \cdots \ud y_{n-1}| = r^{n-1}|\ud r \ud y_1 \cdots \ud y_{n-1}|$ locally with $x$ the boundary defining function and $y_j$'s coordinates for the boundary. For $s > 0$, we define $H^{s, r}_\scop(\br{\R^n})$ to consist functions $u$ such that $Au \in L^2_\scop(\br{\R^n})$ for all $A \in \Psi^{k, r}_\scop(\br{\R^n})$ with $k \leq s$. It is then easy to check the equivalence $H^{s, r}_\scop(\br{\R^n}) = H^{s, r}(\R^n)$. 

These can be generalized to the compact manifold $M$ with boundary.
\begin{definition}[{\cite[Definition 5.10]{Vasynotes}}]\label{defn:Psi-sc-on-M}
    $\Psi_{\mathrm{sc}}^{m, \ell}(M)$ consists of continuous linear maps $A: \dot{\mathcal{C}}^{\infty}(M) \rightarrow \dot{\mathcal{C}}^{\infty}(M)$, whose Schwartz kernel is in $\dot{\mathcal{C}}^{\infty}$ away from the diagonal in $M \times M$ and with the property that if $U$ is a coordinate chart with $\Phi: U \rightarrow \tilde{U} \subseteq \overline{\mathbb{R}^n}$ a diffeomorphism then for $\chi \in \mathcal{C}_c^{\infty}(U),\left(\Phi^{-1}\right)^* \chi A \chi \Phi^* \in \Psi^{m, \ell}_\scop(\br{\R^n})$.
\end{definition}

\begin{proposition}[{\cite[Proposition 5.22]{Vasynotes}}]\label{prop:sc-ell-reg}
    Any elliptic $A \in \Psi_{\scop}^{m, \ell}(M)$ is Fredholm as a map $H_{\mathrm{sc}}^{s, r}(M) \to H_{\mathrm{sc}}^{s-m, r-\ell}(M)$ for all $m, \ell, s, r \in \R$, i.e., it has closed range, finite dimensional nullspace and the range has finite codimension. Further, the nullspace is a subspace of $\dot{\mathcal{C}}^{\infty}(M)$, while the annihilator of the range in $H_{\mathrm{sc}}^{s-m, r-\ell}(M)$ in the dual space $H_{\mathrm{sc}}^{-s+m,-r+\ell}(M)$ is also in $\dot{\mathcal{C}}^{\infty}(M)$. Correspondingly, the nullspace of $A$ as well as the annihilator of its range is independent of $r, s$; if $A$ is invertible for one value of $r, s$, then it is invertible for all.
\end{proposition}

The symbols and Sobolev spaces introduced above can be generalized to permit variable orders/decay; see \cite[Sections 5.3.2, 5.3.9]{Vasynotes}.

We also recall the semiclassical scattering calculus in this setting; see \cite[\S 5.5.2]{Vasynotes}.
In particular, we have the following semiclassical elliptic estimate:
\begin{proposition}[{\cite[Theorem 5.6]{Vasynotes}}]\label{prop:ell-semi}
    Suppose $P_h$ is semiclassically elliptic in $\Psi_h^{m,\ell}(M)$, then we have
    \begin{equation*}
        \|u\|_{H_h^{s,r}}\leq C\|P_h u\|_{H_h^{s-m,r-\ell}}.
    \end{equation*}
\end{proposition}

Now we specialize to our domain of interest $\calM$. The compactified space $\br{\calM}$ is equipped with a global boundary defining function $\rho$ such that $\rho = r_\bop$ near the $N$ horizons and $\rho = r^{-1}$ for $r \gg 1$. We now generalize Definition~\ref{defn:Psi-sc-on-M} to our specific case.

\begin{definition}
    $\Psi_{sc}^{m, (\ell_\infty, \ell_1, \ell_2, \ldots, \ell_N)}(\br{\calM})$ consists of continuous linear maps $A: \dot{\mathcal{C}}^{\infty}(\br{\calM}) \rightarrow \dot{\mathcal{C}}^{\infty}(\br{\calM})$, whose Schwartz kernel is in $\dot{\mathcal{C}}^{\infty}$ away from the diagonal in $\br{\calM} \times \br{\calM}$ and with the property that for an open covering $U_\infty \cup (\cup_{j = 1}^N U_j) = \br{\calM}$ with $U_{\infty}$ around infinity and $U_j$ around $j$'s extremal horizon such that $\Phi_{\bullet}: U_\bullet \rightarrow \tilde{U_\bullet} \subseteq \overline{\mathbb{R}^n}$ a diffeomorphism then for $\chi \in \mathcal{C}_c^{\infty}(U_\bullet),\left(\Phi^{-1}\right)^* \chi A \chi \Phi^* \in \Psi^{m, \ell_\bullet}_\scop(\br{\R^n})$, where $\bullet \in \set{\infty, 1, 2, \ldots, N}$. The independence of choice of coordinate patches can be proved similar to the standard case.
\end{definition}
\begin{remark}
    Here, $\ell_\infty, \ell_1, \ell_2, \ldots, \ell_N$ denote the weight near each boundary of $\br{\calM}$ correspondingly. Equivalently, one can define using the variable coefficient version of symbols. For each coordinate chart $U$ with $\Phi: U \rightarrow \tilde{U} \subseteq \overline{\mathbb{R}^n}$ a diffeomorphism then for $\chi \in \mathcal{C}_c^{\infty}(U),\left(\Phi^{-1}\right)^* \chi A \chi \Phi^* \in \Psi^{m, \ell}_\scop(\br{\R^n})$, where $\ell$ is a smooth function such that \[
    \ell = \begin{cases}
        \ell_j, \qquad \hbox{ for } r_{\bop_j} \ll 1, \\
        \ell_\infty, \qquad \hbox{ for } r \gg 1.
    \end{cases}
    \]
\end{remark}
The definitions of scattering pseudodifferential operators
and weighted scattering Sobolev spaces,
\[
\Psi_\scop^{m,(\ell_\infty,\ell_1,\ldots,\ell_N)}(\br{\calM}),
\qquad H_\scop^{s,(r_\infty,r_1,\ldots,r_N)}(\br{\calM}),
\]
as well as ellipticity, can be generalized analogously
using the charts described above.

\subsection{Semiclassical Weyl's law}
We recall Weyl's law for eigenvalues \cite[Theorem 10.1.1]{Ivrii98}.
\begin{proposition}\label{prop:weyl}
    Let $n\geq 2$, $V\in C^{\infty}(\overline{\Omega};\bbR)$, and $P_h=-h^2\Delta_g+V(x)$ on a bounded domain $\Omega\subseteq \mathbb{R}^n$ with smooth boundary, equipped with Riemannian metric $g$. Let $\lambda_1\leq\lambda_2\leq \cdots$ be the Dirichlet eigenvalues.  Then the counting function $N(\lambda):=\#\{\lambda_j\leq \lambda\}$ with multiplicity satisfies
    \begin{equation}
        N(\lambda)=\frac{1}{(2\pi h)^n}\int_{|\xi|_g^2+V\leq \lambda} \ud x\ud \xi + O(h^{1-n}).
    \end{equation}
\end{proposition}

In particular, we will apply Proposition~\ref{prop:weyl} in the following setting. Equipping a domain $\Omg \subseteq \R^2_{\rho, z}$ ($\p \Omg \in C^\infty$) with $g = U^4(\ud \rho^2 + \ud z^2)$. Then $U^{-4}\lap_{\rho, z} = \lap_g$ and hence the eigenvalue problem of the operator \[
    - U^{-4} h^2 \lap_{\rho, z} + V(\rho, z) = - h^2 \lap_{g} + V(\rho, z)
\]
satisfies the following Weyl's law \begin{align*}
    N(\lmb) &= \frac{1}{2\pi h^2} \int_{\Omg}\int_0^{\sqrt{U^4(\lmb - V)_+}}s\, \ud s\, \ud \rho\, \ud z + O(h^{-1}) \\
    &= \frac{1}{4\pi h^2} \int_{\Omg} U^4(\rho, z)(\lmb - V(\rho, z))_+\, \ud \rho\, \ud z + O(h^{-1}).
\end{align*}

\subsection{Agmon distance}
The following will be used in Section~\ref{sec:quasimode}.
\begin{definition}
    The Agmon distance between two points $p_1, p_2 \in \R_{\rho> 0} \times \R_z$ associated to the energy level $E$ and potential $V$ is given by \[
        d_E(p_1, p_2) := \inf_{\gmm\, \mathrm{ piecewise }\, C^1, \gmm(0) = p_1, \gmm(1) = p_2} \int_0^1 \sqrt{U(\gmm(t))^4(V(\gmm(t)) - E)_+} |\gmm'(t)| \chi_{V \geq E} \, \ud t,
    \]
    where $(V(\gamma(t))-E)_+ \coloneqq \max\{V(\gamma(t))-E,0\}$.
    Moreover, we define the Agmon distance from the classically allowed region \[
        d_E(p_1) := \inf_{p_2 : V(p_2) \leq E} d_E(p_1, p_2).
    \]
\end{definition}

\begin{lemma}\label{lem:Agmon-gradient}
    The Agmon distance satisfies the pointwise gradient estimate \[
        |\nabla_{p_1} d_E(p_1, p_2)|^2 \leq \max \set{U^4(p_1)(V(p_1) - E), 0}
    \]
    for almost every $p_1, p_2$.
\end{lemma}
\begin{proof}
    For any infinitesimal displacement $v$, we have \[
        |d_E (p_1 + v, p_2) - d_E(p_1, p_2)| \leq d_E(p_1 + v, p_1) \leq |v|\sup_{\td{p} \in B(p_1, |v|)} (\sqrt{U^4(\td p)(V(\td p) - E)} \chi_{V(\td p) \geq E}).
    \]
    Taking $|v| \to 0$ concludes the proof.
\end{proof}

\section{Meromorphic continuation and QNM: Proof of Theorem~\ref{thm:meromorphic-ext}}\label{sec:meromorphic-cont}
The main purpose of this section is to prove Theorem~\ref{thm:meromorphic-ext}. Our implementation of the method of complex scaling in Sections~\ref{subsec:complex-scaled-op}~-~\ref{subsec:scattering-ell} is mostly based on the exposition from \cite[Sections 3.2, 3.7]{stucker2024} on subextremal Kerr spacetime (see also \cite{GajicWarnick24} for a parallel treatment). Owing to the extremality and particular structure of the metric, we select cutoff functions whose supports are away from both the horizons and infinity, and these regions are regarded as asymptotic ends using scattering calculus.

\subsection{The complex scaled operator}\label{subsec:complex-scaled-op}
In the following, we introduce several parameters to be determined such that \begin{equation}\label{eq:R012-relation}
    R_{\bop, 2} < R_{\bop, 1} < R_{\bop, 0}\ll 1, \qquad 1 \ll R_{\infty, 0} < R_{\infty, 1} < R_{\infty, 2}.
\end{equation}
\begin{definition}
    Given $R_{\infty, 1} > \max_{j = 1}^N |\bfr_j|$, $0 < R_{\bop, 1} < \frac12 \min_{j \neq \ell} |\bfr_j - \bfr_\ell|$ and $\eps > 0$. For $\bt \in (-\pi, \pi)$, we define the complex scaling map $F_\bt: \calM (\subseteq \R^3) \to \bbC^3$ by  \begin{equation}\label{eq:Fbt-defn}
        F_\bt(x) = \left\{ \begin{split}
            &\bfr + e^{i\bt\psi_\bop(\log |x - \bfr|)} (x - \bfr), &r_{\bop} = |x - \bfr| < R_{\bop, 1},\\
            &e^{-i\bt\psi(\log |x|)}x, &r = |x| > R_{\infty, 1},\\
            &x, &\hbox{otherwise},
        \end{split}\right.
    \end{equation}
    where $\psi_\bop, \psi \in C^\infty(\R; \R)$ are chosen to satisfy the following for some $R_{\bop, 2} = R_{\bop, 2}(\eps, R_{\bop, 1}) < R_{\bop, 1}$ and $R_{\infty, 2} = R_{\infty, 2}(\eps, R_{\infty, 1}) > R_{\infty, 1}$: \begin{itemize}
        \item $\psi_\bop(\R), \psi(\R) \subseteq [0, 1]$;
        \item $|\psi_\bop'(s)| + |\psi'(s)| \leq \eps$; 
        \item $\psi_\bop = 0$ on $(\log(R_{\bop, 1}), \infty)$ and $\psi = 0$ on $(-\infty, \log(R_{\infty, 1}))$;
        \item $\psi = 1$ on $(\log(R_{\infty, 2}), \infty)$ and $\psi_\bop = 1$ on $(-\infty, \log(R_{\bop, 2}))$.
    \end{itemize}
\end{definition}
\begin{remark}
    Note that $\eps$, $R_{\bullet, 0}$ and $R_{\bullet, 1}$ ($\bullet = \bop, \infty$) are still flexible here, which will be determined according to \eqref{eq:ord-symbol-inside}, Lemma~\ref{lem:deformation} and Lemma~\ref{lem:cutoff-res-bt-0}, respectively. We also note that it is essential to have compatible sign conventions in the phase of \eqref{eq:Fbt-defn} near infinity/horizons; see Remark~\ref{rmk:compatible-sign-bt} for a further discussion.
\end{remark}

\begin{lemma}
    $\calM_\bt := F_\bt(\calM)$ is a maximally totally real submanifold of $\bbC^3$.
\end{lemma}
\begin{proof}
    Since $\bfr_j$'s are distinct, by choosing $R_{\bop, 1} \ll 1$ and $R_{\infty, 1} \gg 1$, we can ensure that the sets $F_\bt(B(\bfr_j,R_{\bop, 1}))$ are pairwise disjoint for different $j$. Thus, the injectivity of $F_\bt$ follows from the injectivity of its restriction to each region, together with the disjointness of their images.
    
    For $z = F_\bt(x)$, we compute the Jacobian \begin{equation}\label{eq:Jacobian-udF}
        (\ud F_\bt(x))_{jk} = \tfrac{\p z^j}{\p x^k} = \left\{\begin{split}
            &e^{i\bt\psi_\bop(\log |x - \bfr|)}\left(\dlt_{jk} + i\frac{\bt(x^j - \bfr^j)(x^k - \bfr^k)}{r_\bop^2}  \psi_\bop'(\log r_\bop)\right), \quad &r_\bop < R_{\bop, 1}, \\
            &e^{-i\bt\psi(\log |x|)}\left(\dlt_{jk} - i\frac{\bt x^j x^k}{r^2}  \psi'(\log r)\right), \quad &r > R_{\infty, 1}, \\
            &\dlt_{jk}, \quad &\hbox{ otherwise}.
        \end{split}\right.
    \end{equation}
    It is easy to see that for any $v \in \bbC^3 \setminus \set{0}$, $\br v \cdot \ud F_\bt(x) v \neq 0$ and hence $\ud F_\bt: \bbC^3 \to \bbC^3$ is injective. Thus, thanks to \eqref{eq:totally-real-udF}, the lemma follows.
\end{proof}

Note that $F_\bt(x)$ agrees with $x$ in a region away from the horizons and infinity, namely in $x \in B(0, R_{\infty, 1}) \setminus \cup_{j = 1}^N B(\bfr_j, R_{\bop, 1})$. $\calM_\bt$ is a deformation of $\calM$ to $\bbC^3$ outside this region. We now want to apply complex scaling to $\td P(\lmb) = -\lap - \lmb^2 U^4(x)$. We take an open subset of $\calM$ \begin{equation}\label{eq:defn-Omg}
    \Omg := \big(\calM \setminus \br{B(0, R_{\infty, 0})}\big) \cup \big(\cup_{j = 1}^N B(\bfr_j, R_{\bop, 0}) \cap \calM\big) =: \Omg_\infty \cup (\cup_{j = 1}^N \Omg_j) 
\end{equation}
and look for an open subset $\calU \subseteq \bbC^3$ such that \begin{equation}\label{eq:calU-requirement}
    \cup_{\bt \in [0, \tfrac{\pi}{2} - \dlt)} F_\bt(\Omg) \subseteq \calU 
\end{equation}
for some $\dlt > 0$ and the coefficients of $\td P(\lmb)$ extend holomorphically on $\calU$. Indeed, such an $\calU$ exists; see Lemma~\ref{lem:calU-set}. 

\begin{lemma}\label{lem:calU-set}
    Given any $\dlt \in (0, \frac{\pi}{4})$, there exist $R_{\infty, 0} \gg 1$ and $R_{\bop, 0} \ll 1$ such that 
    \begin{equation}\label{eq:defn-calU}
        \begin{split}
            \calU &:= \left\{z \in \bbC^3 : |(z^1)^2 + (z^2)^2 + (z^3)^2| > R_{\infty, 0}^2,\right. \\
            &\qquad\qquad\qquad\qquad\qquad\qquad\qquad \left.\hbox{ or } \exists j,\, 0 < |(z^1 - \bfr_j^1)^2 + (z^2 - \bfr_j^2)^2 + (z^3 - \bfr_j^3)^2| < R_{\bop, 0}^2 \right\}  \\
            &\,\peq\cap \left\{z \in \bbC^3 : \arg((z^1 - \bfr_j^1)^2 + (z^2 - \bfr_j^2)^2 + (z^3 - \bfr_j^3)^2) \in (-\pi, \pi) \hbox{ for all } j = 1, 2, \ldots, N\right\}
        \end{split}
    \end{equation}
    is an open subset of $\bbC^3$ satisfying \eqref{eq:calU-requirement}, on which $\td{P}(\lmb)$ defines a differential operator with analytic coefficients. 
\end{lemma}
\begin{proof}
    It is obvious that $\Omg \subseteq \calU$ and $U(x)$ can be extended holomorphically to $\calU$ as \[
        U(z) = 1 + \sum_{j = 1}^N \frac{M_j}{\sqrt{(z^1 - \bfr_j^1)^2 + (z^2 - \bfr_j^2)^2 + (z^3 - \bfr_j^3)^2}}
    \]
    where $z = (z^1, z^2, z^3) \in \bbC^3$ should not be confused with the notation $z$ in the cylindrical coordinates. Indeed, the first restriction in \eqref{eq:defn-calU} ensures the terms under the square root are nonzero while the second condition selects a holomorphic branch. Now we verify \eqref{eq:calU-requirement}. It suffices to verify the condition on the arguments in \eqref{eq:defn-calU}.

    {\noindent \it Step 1: }
    For $x \in \Omg_J$, $J \in \set{1, \ldots, N}$, we express \[
         (z^1, z^2, z^3) = F_\bt(x) =  (\bfr_J^1 + e^{i\bt\psi_\bop} (x^1 - \bfr_J^1), \bfr_J^2 + e^{i\bt\psi_\bop} (x^2 - \bfr_J^2), \bfr_J^3 + e^{i\bt\psi_\bop} (x^3 - \bfr_J^3)).
    \]
    It is obvious that \[
        \arg((z^1 - \bfr_J^1)^2 + (z^2 - \bfr_J^2)^2 + (z^3 - \bfr_J^3)^2) \in [0, 2\bt] \in [0, \pi).
    \]
    On the other hand, for any $j \neq J$, we write 
    \begin{align*}
         &\peq\Re\sum_{k = 1}^3 (z^k - \bfr_j^k)^2 
         = \sum_{k = 1}^3 \Re \big((\bfr_J^k - \bfr_j^k) + e^{i\bt\psi_\bop}(x^k - \bfr_J^k)\big)^2 \\
         &= \sum_{k = 1}^3 (\bfr_J^k - \bfr_j^k)^2 + \cos(2\bt\psi_\bop) \sum_{k = 1}^3(x^k - \bfr_J^k)^2 + 2\cos(\bt\psi_\bop)\sum_{k = 1}^3(\bfr_J^k - \bfr_j^k)(x^k - \bfr_J^k),
    \end{align*}
    which is positive by taking $|x^k - \bfr_J^k| < R_{\bop, 0}$ sufficiently small compared to $\sum_{k = 1}^3 (\bfr_J^k - \bfr_j^k)^2$. Hence, $\arg(\sum_{k = 1}^3 (z^k - \bfr_j^k)^2) \in (-\pi, \pi)$.

    
    {\noindent \it Step 2: }
    For $x \in \Omg_\infty$, with an abuse of notation, we represent \[
         (z^1, z^2, z^3) = F_\bt(x) = (e^{-i\bt\psi} x^1, e^{-i\bt\psi} x^2, e^{-i\bt\psi} x^3).
    \]
    Note that \[
        \sum_{k = 1}^3 (z^k - \bfr_j^k)^2 
        = e^{-2i\bt\psi}\sum_{k = 1}^3 (x^k - e^{i\bt\psi}\bfr_j^k)^2.
    \]
    It is obvious that \[
        \Re \sum_{k = 1}^3 (x^k - e^{i\bt\psi}\bfr_j^k)^2 \geq \frac12 r^2, \qquad |\Im \sum_{k = 1}^3 (x^k - e^{i\bt\psi}\bfr_j^k)^2| \aleq r
    \]
    for $r>R_{\infty, 0} \gg 1$. Hence, by choosing $R_{\infty, 0}$ sufficiently large so that $|\Arg(\sum_{k = 1}^3 (x^k - e^{i\bt\psi}\bfr_j^k)^2)| < \dlt$, \[
        \arg(\sum_{k = 1}^3 (z^k - \bfr_j^k)^2) \in (-2\bt \psi - \dlt, -2\bt\psi + \dlt) \subseteq (-\pi + \dlt, \dlt),
    \]
    which finishes the proof.
\end{proof}
\begin{remark}\label{rmk:compatible-sign-bt}
    \begin{enumerate}[leftmargin = *]
        \item We note that incompatibility of sign conventions in the phase would not lead to nonexistence of such an $\calU$ in the preceding lemma. However, it shall become clear in \eqref{eq:scattering-symbol} and the corresponding choice of $\Lmb_\bt$ that the opposite sign convention in \eqref{eq:Fbt-defn} near horizons and infinity is crucial to reveal the scattering ellipticity in both regions. 
        \item On the other hand, for $\bt \in [-\frac{\pi}{2} + \dlt, 0)$, the corresponding definition of $\calU$ still works but we remark that choosing the same $\calU$ for each subrange of $\bt$ is not necessary.  Since the modification is relatively direct, we will only work out the case with $\bt$ positive unless otherwise specified.
    \end{enumerate}
\end{remark}

Thanks to Lemma~\ref{lem:calU-set}, $-\sum_{j = 1}^3 \p_{z^j}^2 - \lmb^2 U^4(z)$ is a holomorphic differential operator on $\calU$ for each $\bt$ in the range considered in \eqref{eq:calU-requirement}. Therefore, applying Lemma~\ref{lem:unique-ext-op-on-TRS}, we know that there exists a unique differential operator $\td P_{\calM_\bt}(\lmb)$ such that for every $u \in C^\infty(F_\bt(\Omg))$, we have \begin{equation}\label{eq:tdPM3bt}
    \td P_{\calM_\bt}(\lmb) u = \left(-\sum_{j = 1}^3 \p_{z^j}^2 \td u - \lmb^2 U^4(z) \td u\right)\rst_{z \in F_\bt(\Omg)},
\end{equation}
where $\td u$ is an almost analytic extension of $u$ to $\bbC^3$. 

\begin{definition}
    For each $\bt$ in the range considered in \eqref{eq:calU-requirement}, we define a complex scaled operator $\td P_\bt(\lmb)$ on $\calM_\bt = F_\bt(\calM)$ by \begin{equation}\label{eq:defn-tdPbt}
        \td P_\bt(\lmb) := \begin{cases}
            \td P_{\calM_\bt}(\lmb) &\hbox{ on } F_\bt(\Omg), \\
            \td P(\lmb) &\hbox{ on } \calM_\bt \setminus F_\bt(\Omg) \subseteq \calM,
        \end{cases}
    \end{equation}
    where $\td P_{\calM_\bt}(\lmb)$ is defined as \eqref{eq:tdPM3bt}.
\end{definition}

According to \eqref{eq:R012-relation}, we observe that \eqref{eq:defn-tdPbt} admits a smooth transition and the expression \eqref{eq:tdPM3bt} turns out to be valid on the entire domain. More precisely, the operator takes the form (in the coordinate $F_\bt^{-1}: \calM_\bt \to \calM$):
\begin{equation}\label{eq:tdP-bt-lmb}
    \begin{split}
        (\td P_\bt(\lmb) u)(F_\bt(x)) &= -\left(\sum_{j = 1}^3 \p_{z^j}^2 u(z) + \lmb^2 U^4(z) u(z)\right)\rst_{z = F_\bt(x)} \\ 
        &= -({ }^{\top} \ud F_\bt(x)^{-1} \cdot \p_x)^2 (u \circ F_\bt)(x) - \lmb^2 U^4(F_\bt(x)) u(F_\bt(x)),
    \end{split}
\end{equation}
where $\p_{x^j} = (\p_{x^j} F_\bt^k)\p_{z^k}$. 
\begin{remark}\label{rmk:tdPbt-coord-rep}
    We note that the representation of $\td P_\bt(\lmb)$ in the coordinate $F_\bt^{-1}$  \eqref{eq:tdP-bt-lmb} can be viewed as an operator on $\calM$, which is the same as the differential operator \[
        F_\bt^* \circ \td P_\bt(\lmb) \circ (F_\bt^{-1})^* : C^\infty(\calM) \to C^\infty(\calM).
    \]
    In what follows, we adopt a slight abuse of notation by identifying the two operators and, correspondingly, the associated function spaces when no ambiguity arises.
\end{remark}

\subsection{Scattering ellipticity and Fredholm property}\label{subsec:scattering-ell}
Now we explicitly expand \eqref{eq:tdP-bt-lmb} in certain regions, which allows us to examine its scattering ellipticity.
Thanks to Lemma~\ref{lem:calU-set}, we write \[
    \td P_\bt(\lmb) = \left\{
    \begin{split}
    &- e^{2i\bt}\lap - \lmb^2 \left(1 + \sum_{j = 1}^N\frac{e^{i\bt}M_j}{\sqrt{\sum_{\ell = 1}^3 ( x^\ell - e^{i\bt}\bfr_j^\ell)^2}} \right)^4, \quad |x| > R_{\infty, 2}, \\
    &- e^{-2i\bt}\lap - \lmb^2 \left(1 + \sum_{j = 1}^N\frac{e^{-i\bt}M_j}{\sqrt{\sum_{\ell = 1}^3 ( (x^\ell - \bfr^\ell_J) - e^{-i\bt}(\bfr_j^\ell - \bfr_J^\ell))^2}} \right)^4, \quad \begin{split}
        &|x - \bfr_J| < R_{\bop, 2},\\ &1 \leq J \leq N.
    \end{split}
    \end{split}\right.
\]
Therefore, by using boundary defining functions $\frkr := 1/r$, $r_{\bop_J}$ ($1 \leq J \leq N$), respectively, in corresponding regions, the operator can be expressed as \begin{equation}\label{eq:scattering-symbol}
    \begin{cases}
    -e^{2i\bt} \left((\frkr^2\p_{\frkr})^2 - 2\frkr^3\p_{\frkr} + \frkr^2 \Dlt_{\bbS^2}\right) - \lmb^2(1 + O(\frkr))^4, \hbox{ in } \set{r > R_{\infty, 2}}, \\
    -r_{\bop_J}^{-4}\left(e^{-2i\bt}\big((r_{\bop_J}^2\p_{r_{\bop_J}})^2 + r_{\bop_J}^2 \Dlt_{\bbS^2}\big) + \lmb^2 (e^{-i\bt}M_J + O(r_{\bop_J}))^4\right), \hbox{ in } \set{r_{\bop_J} < R_{\bop, 2}}.
\end{cases} 
\end{equation}
Set \[
    \Lmb_\bt := \set{\lmb \in \bbC \setminus \set{0}: \arg(\lmb) \in (\bt, \pi + \bt)}.
\]
It is obvious that, for $\lmb \in \Lmb_\bt$, $\chi_{\infty}\td P_\bt(\lmb)\chi_{\infty}$ (resp. $\chi_\bop r_\bop^4\td P_\bt(\lmb)\chi_\bop$) is in $\Psi^{2, 0}_\scop(\br{\calM_\bt})$, where $\chi_\infty$ (resp. $\chi_\bop$) is localized to $\set{r > R_{\infty, 2}}$ (resp. $\set{r_\bop < R_{\bop, 2}}$). 

On the other hand, in order to ensure the ordinary ellipticity of $\td P_\bt(\lmb)$ in their complementary regions, we require $\eps \ll 1$. Namely, by combining \eqref{eq:Jacobian-udF} and \eqref{eq:tdP-bt-lmb}, we directly read off the ordinary principal symbol of the operator $\td P_\bt(\lmb)$: \begin{equation}\label{eq:ord-symbol-inside}
    \begin{cases}
        e^{-2i\bt\psi_\bop} \sum_{j = 1}^3 \left( \sum_{k = 1}^3 \left(\left(\dlt + i\bt \frac{(x - \bfr)\otimes (x - \bfr)}{|x - \bfr|^2} \psi_\bop'\right)^{-1}\right)_{jk}\xi_k\right)^2 \\
        \qquad {}= e^{-2i\bt\psi_\bop}(1 + O(\eps))|\xi|^2, \qquad R_{\bop, 2} \leq r_\bop \leq R_{\bop, 1}, \\
        e^{2i\bt\psi} \sum_{j = 1}^3 \left( \sum_{k = 1}^3 \left(\left(\dlt - i\bt \frac{x \otimes x}{|x|^2} \psi'\right)^{-1}\right)_{jk}\xi_k\right)^2 \\
        \qquad {}= e^{2i\bt\psi}(1 + O(\eps))|\xi|^2, \qquad R_{\infty, 1} \leq r \leq R_{\infty, 2}, \\
        |\xi|^2, \qquad \hbox{ otherwise}.
    \end{cases}
\end{equation}
Thus, we conclude that 
\[
    \td P_\bt(\lmb) \in \Psi_\scop^{2, (0, 4, 4, \ldots, 4)}(\br{\calM_\bt}), \qquad \forall \lmb \in \Lmb_\bt,
\]
where the space is introduced in Section~\ref{subsec:sc}. Therefore, for all $\lmb \in \Lmb_\bt$, the operator \begin{equation}\label{eq:tdPbt-mapping-prop}
    \lmb \in \Lmb_\bt \mapsto \td P_\bt(\lmb): H^{s, (r_\infty, r_1, \ldots, r_N)}_\scop(\br{\calM_\bt}) \to H^{s - 2, (r_\infty, r_1 - 4, \ldots, r_N - 4)}_\scop(\br{\calM_\bt})
\end{equation}
is an analytic Fredholm operator. We read off from \eqref{eq:Rlmb-Lmb0} that $\td P_0(\lmb)$ is invertible for $\Im \lmb > 0$ for $(s, r_\infty, r_1, \ldots, r_N) = (2, 0, -1, \ldots, -1)$, where $L^2(\calM; U^4\, \ud x) = H^{0, (0, -1, \ldots, -1)}_\scop(\br{\calM})$. By Proposition~\ref{prop:sc-ell-reg}, if the operator is invertible for a single tuple $(s, r_\infty, r_1, \ldots, r_N)$, then it is invertible for all such tuples. To define $\td P_\bt(\lmb)^{-1}$ suitably, it suffices to know that \eqref{eq:tdPbt-mapping-prop} has index $0$ for some $(s, r_\infty, r_1, \ldots, r_N)$. Therefore, we have the following results in order.

\begin{lemma}\label{lem:kernel-relation}
    Choosing $R_{\bop, 0} \ll 1$ and $R_{\infty, 0} \gg 1$, we define $\td P_\bt(\lmb)$ according to Section~\ref{subsec:complex-scaled-op}.
    Let $\bt_1, \bt_2 \in [0, \frac{\pi}{2})$ or $\bt_1, \bt_2 \in (-\frac{\pi}{2}, 0]$. Let $\lmb \in \Lmb_{\bt_1} \cap \Lmb_{\bt_2}$, then \begin{equation}\label{eq:dim-ker-tdPbt}
        \dim(\ker \td P_{\bt_1}(\lmb)) = \dim(\ker \td P_{\bt_2}(\lmb)). 
    \end{equation}
\end{lemma}
\begin{proof}
    We first note that $\ker \td P_{\bt}(\lmb)$ is independent of choices of weights $s$ and $(r_\infty, r_1, \ldots, r_N)$ in \eqref{eq:tdPbt-mapping-prop} since the kernel elements are Schwartz thanks to scattering elliptic regularity (Proposition~\ref{prop:sc-ell-reg}). 
    
    It suffices to prove \eqref{eq:dim-ker-tdPbt} for $|\bt_1 - \bt_2| \ll 1$. For any $u_1 \in \ker \td P_{\bt_1}(\lmb)$, we have $u_1 \in \dot C^\infty(\br{\calM})$.

    {\it Step 1: Construction of $u_2$. }
    The idea is to obtain a family of analytic extensions by applying Lemma~\ref{lem:deformation} in a family of compact subsets of $\calU$, which gives rise to $u_2 \in C^\infty(\calM_{\bt_2})$ such that $\td P_{\bt_2}(\lmb)u_2 = 0$. Here, $u_2$ is obtained through certain analytic extension, which is only necessary in $\Omg$; see \eqref{eq:defn-Omg}. Indeed, we note that $u_2$ is expected to agree with $u_1$ on where $F_\bt(x) = x$ (see \eqref{eq:Fbt-defn}) and decompose its complement into dyadic pieces: \begin{align*}
        &\calM \setminus \br{B(0, R_{\infty, 1})} = \cup_{k \geq 0} \set{ 2^k R_{\infty, 1} \leq r \leq 2^{k+1} R_{\infty, 1} } =: \cup_{k \geq 0} A_{\infty, k}, \\
        &B(\bfr_j, R_{\bop, 1}) = \cup_{k \geq 0} \set{2^{-k-1} R_{\bop, 1} \leq r_{\bop_j} \leq 2^{-k} R_{\bop, 1}} =: \cup_{k \geq 0} A_{\bop_j, k}, \qquad j = 1,2,\ldots, N.
    \end{align*}
    Moreover, we define finitely overlapping enlargement $\td A_{\bullet, k} \supseteq \wh A_{\bullet, k} \supseteq A_{\bullet, k}$ such that they all lie in $\Omg$, where $\bullet \in \set{\infty, \bop_1, \bop_2, \ldots, \bop_N}$ and $\Omg = \cup_{\bullet} \cup_{k \geq 0} \td A_{\bullet, k}$. For simplicity, we also ensure that $\td A_{\bullet, k} = 2^{\sgn(\bullet) k} \td A_{\bullet, 0}$ and $\wh A_{\bullet, k} = 2^{\sgn(\bullet)k} \wh A_{\bullet, 0}$ with $\sgn(\infty) = 1$ and $\sgn(\bop_J) = -1$, where the dilation is with respect to $r$ and $r_\bop$ for $\bullet = \infty, \bop$, respectively.

    We now modify \eqref{eq:Fbt-defn} to obtain a family of maps $F_{\bt_1, \bt_2}^{s,k}$ with a family of operators $\td P_{\bt_1, \bt_2}^{s, k}$ associated:
    \begin{equation}
        F_{\bt_1, \bt_2}^{s,k}(x) = \left\{ \begin{split}
            &\bfr + e^{i(\bt_1 + s\chi_{\td A_{\bop, k}}(r_\bop)(\bt_2 - \bt_1))\psi_\bop(\log r_\bop)} (x - \bfr), &r_{\bop} < R_{\bop, 1},\\
            &e^{-i(\bt_1 + s\chi_{\td A_{\infty, k}}(r)(\bt_2 - \bt_1))\psi(\log r)}x, &r > R_{\infty, 1},\\
            &x, &\hbox{otherwise},
        \end{split}\right.
    \end{equation}
    where $\chi_{\td A_{\bullet, k}}$ is supported on $\wh A_{\bullet, k} \subseteq \td A_{\bullet, k}$ and is equal to one on $A_{\bullet, k}$. 
    Fixing $k$, we apply Lemma~\ref{lem:deformation} to $\cup_{s \in [0, 1]} F_{\bt_1, \bt_2}^{s,k}(\cup_\bullet \td A_{\bullet, k})$, which gives $u_{\bt_1, \bt_2, k}$ such that $\td P_{\bt_1, \bt_2}^{s, k}(\lmb)u_{\bt_1, \bt_2, k} = 0$. By elliptic regularity, $u_{\bt_1, \bt_2, k} \in C^\infty$. We define $u_2$ as \[
        u_2 := \left\{ \begin{split}
            u_{\bt_1, \bt_2, k}, \hbox{ on }& F_{\bt_2}(A_{\bullet, k}), \\
            u_1, \hbox{ on }& B(0, R_{\infty, 1}) \cap (\calM \setminus \cup_j \br{B(0, R_{\bop_j, 1})}),
        \end{split}\right.
    \]
    where $A_{\bullet, k} \subseteq \wh A_{\bullet, k} \subseteq \td A_{\bullet, k}$. Note that $u_2$ is well-defined since the assembling procedure above is well-defined, i.e., although $F_{\bt_2}(A_{\bullet, k})$ overlaps with each other, the extensions $u_{\bt_1, \bt_2, k}$ obtained from different $k$'s agree on common domains thanks to the fact that two distinct analytic functions cannot agree on an open set (they both equal to $u_1$ in some region). Now it suffices to show that $u_2 \in H^{2, \vec r}_\scop(\br{\calM_{\bt_2}})$ for some $\vec r :=(r_\infty, r_1, \ldots, r_N)$ to ensure $u_2 \in \ker \td P_{\bt_2}(\lmb)$.
    
    {\it Step 2: Reduction to dyadic estimates. }
    We claim that \[
    \|u_2\|_{H^{2, (0,0,\ldots,0)}_\scop(\br{\calM_{\bt_2}})}
    \aleq \|u_1\|_{H^{2, (0,0,\ldots,0)}_\scop(\br{\calM_{\bt_1}})}.\]
     
    It suffices to show that \begin{equation}\label{eq:est-u2-Hsc}
        \|u_2\|_{H^{2, (0,0,\ldots,0)}_\scop(A_{\bullet, k})} \aleq \|u_1\|_{H^{2, (0,0,\ldots,0)}_\scop(\td A_{\bullet, k})}
    \end{equation}

    {\it Step 3: Property of model operators close to variants of $\td P_\bt(\lmb)$.}
    In this step, we first consider \[
        P = -e^{2i\bt}\lap_{\R^3} - M_\bullet^4 \lmb^2, \qquad M_\bullet \in \set{M_1, M_2, \ldots, M_N, M_\infty := 1}
    \]
    where the form of $P$ is motivated by \eqref{eq:scattering-symbol}.
    For $\lmb \in \Lmb_\bt$, $\|v\|_{H^2(\R^3)} \aleq \|Pv\|_{L^2(\R^3)}$.
     By using the cutoff argument in \cite[(29)]{stucker2024} or the semiclassical rescaling argument in \cite[(3.19)]{SjZw91}, \[
        \|v\|_{H^2(2^kV)} \aleq \|Pv\|_{L^2(2^k\td V)} + \|v\|_{L^2(2^k(\td V \setminus V))},    \]
    where $V \subsetneq \td V \subseteq \R^3$ and the implicit constant is independent of $k$. Transferring to the scattering setting near infinity (i.e., when $k \gg 1$), \[
        \|v\|_{H^{2, 0}_\scop(2^{-k}V)} \aleq \|-e^{2i\bt}((\rho^2\p_\rho)^2 + \rho^2\lap_{\bbS^2})v - M_\bullet^4\lmb^2v\|_{H^{0, 0}_\scop(2^{-k}\td V)} + \|v\|_{H^{0, 0}_\scop(2^{-k}(\td V \setminus V))},
    \]  
    where $V \subsetneq \td V \subseteq \R^3$ and the implicit constant is independent of $k$.

    {\it Step 4: Proof of \eqref{eq:est-u2-Hsc} using intermediate contours between angle $\bt_1, \bt_2$.}
    First, thanks to $R_{\infty, 0} \gg 1$, $R_{\bop, 0} \ll 1$ and the smallness of $|\bt_1 - \bt_2|$, we note the smallness \begin{align*}
         &\|\td P_{\bt_1, \bt_2, k}(\lmb) + e^{2i\bt_1}((\frkr^2\p_\frkr)^2 + \frkr^2\lap_{\bbS^2}) + \lmb^2\|_{H^{2, 0}_\scop(\td A_{\infty, k}) \to L^2_{\scop}(\td A_{\infty, k})} \leq \eps\\ 
         &\|r_\bop^4 e^{4i\bt_1}\td P_{\bt_1, \bt_2, k}(\lmb) + e^{2i\bt_1}((r_\bop^2\p_{r_\bop})^2 + r_\bop^2\lap_{\bbS^2}) + \lmb^2M_j^4\|_{H^{2, 0}_\scop(\td A_{\bop, k}) \to L^2_{\scop}(\td A_{\bop, k})} \leq \eps,
     \end{align*}
     where we abuse notations identifying $A_{\infty, k}$ with its image under radial compactification. 
     
     Applying the final estimates in Step~3 to $u^{\bt_1, \bt_2}_k$ with $\td V = \td A_{\bullet, 0}$ and $V = \wh A_{\bullet, 0}$, \eqref{eq:est-u2-Hsc} follows.
\end{proof}
\begin{remark}
    One needs to take special caution on the range of $\lmb$ such that \eqref{eq:dim-ker-tdPbt} holds when $-\tfrac{\pi}{2} < \bt_1 < 0 < \bt_2 < \tfrac{\pi}{2}$. Indeed, we should assume $\lmb \in \Lmb_{\bt_1} \cap \Lmb_{\bt_2} \cap \Lmb_0$ in this case since one needs to go through $\bt = 0$.
\end{remark}

\begin{lemma}\label{lem:Pbt-inv}
    For any $\bt \in (-\tfrac{\pi}{2}, \tfrac{\pi}{2})$, any tuple $(s, r_\infty, r_1, \ldots, r_N)$, the analytic family of Fredholm operators \eqref{eq:tdPbt-mapping-prop} is of index $0$ and is invertible away from a discrete subset of $\Lmb_\bt$, i.e., we have a meromorphic family of operators \[
        \lmb \in \Lmb_\bt \mapsto \td P_\bt(\lmb)^{-1}: H^{s-2, (r_\infty, r_1-4, \ldots, r_N-4)}_\scop(\br{\calM_\bt}) \to H^{s,(r_\infty, r_1, \ldots, r_N)}_\scop(\br{\calM_\bt})
    \]
    with poles of finite rank.
\end{lemma}
\begin{proof}
    For each $\bt$, in order to apply the analytic Fredholm theorem, we need to make sure $\td P_\bt(\lmb)$ is invertible for at least one $\lmb$. Since $\set{\Im \lmb > 0} \cap \Lmb_\bt \neq \0$, it reduces to show that $\td P_\bt(\lmb)$ is of index $0$ thanks to \eqref{eq:U4-normalized-resolvent}, Lemma~\ref{lem:kernel-relation} and the discussion preceding it.
    This then follows from the local constancy of Fredholm index and continuity of $\td P_\bt(\lmb)$ in $\bt$. We note that even though the function spaces are different for different $\bt$'s, it suffices to consider the coordinate representation $F_\bt^* \td P_\bt(\lmb) (F_\bt^*)^{-1}$ as in Remark~\ref{rmk:tdPbt-coord-rep}. This finishes the proof.
\end{proof}

Now we proceed to establish the meromorphic continuation of $\chi \td{P}(\lambda)^{-1}\chi$. It is an easy corollary of the following lemma.

\begin{lemma}\label{lem:cutoff-res-bt-0}
    Let $\bt_1, \bt_2 \in [0, \tfrac{\pi}{2})$ or $\bt_1, \bt_2 \in (-\tfrac{\pi}{2}, 0]$. Let $\chi \in C^\infty_{\rm{c}}(\calM)$ such that $R_{\bullet, 1}$ ($\bullet = \bop, \infty$) is chosen to satisfy $\supp \chi \subseteq \set{r < R_{\infty, 1}} \cap \set{r_\bop > R_{\bop, 1}}$. Then for any $\lmb \in \Lmb_{\bt_1} \cap \Lmb_{\bt_2}$ with the property that $\ker \td P_\bt(\lmb)$ is trivial for all $\bt$ such that $\lmb \in \Lmb_\bt$, we have \[
        \chi \td P_{\bt_1}(\lmb)^{-1} \chi = \chi \td P_{\bt_2}(\lmb)^{-1} \chi
    \]
    as operators on $L^2(\calM) \to L^2(\calM)$.
\end{lemma}
\begin{proof}
    Given $f \in L^2(\calM)$, $\chi f \in H^{0, (0, -4,\ldots, -4)}_\scop(\br{\calM})$. Let \[
        u_1 := \td P_{\bt_1}(\lmb)^{-1} \chi f, \qquad u_2 := \td P_{\bt_2}(\lmb)^{-1} \chi f
    \]
    and it suffices to prove $\chi u_1 = \chi u_2$. 
    
    On the other hand, we observe that $\td P_{\bt_1} u_1 = \chi f$, where the right hand side vanishes identically in the region in which the complex scaling takes place. Hence, using the same argument as in the proof of Lemma~\ref{lem:kernel-relation}, we could obtain $\td u_2 \in H^{2, (0,0,\ldots,0)}_\scop(\br{\calM_{\bt_2}})$ satisfying $\td u_2 = u_1$ on $\supp \chi$ and $\td P_{\bt_2}(\lmb) \td u_2 = \chi f$. Indeed, this is because we only perform the analytic extension in a family of compact sets outside $\supp \chi$, i.e., where complex scaling takes place.
    
    By invertibility, $\td u_2 = u_2$. Thus $\chi u_1 = \chi \td u_2 = \chi u_2$ and the proof is complete.
\end{proof}

Theorem~\ref{thm:meromorphic-ext} now follows.

\subsection{Proof of Theorem~\ref{thm:meromorphic-ext}}

\begin{proof}[Proof of Theorem~\ref{thm:meromorphic-ext}]
    For each $\bt \in (-\tfrac{\pi}{2}, \frac{\pi}{2})$ and $\chi \in C^\infty_{\rm{c}}(\calM)$, we choose an associated $R_{\bullet, 1}$ by Lemma~\ref{lem:cutoff-res-bt-0} such that \[
        \chi \td P_\bt(\lmb)^{-1} \chi = \chi \td P_0(\lmb)^{-1} \chi, \hbox{  on } \Lmb_\bt \cap \Lmb_0.
    \]
    Thus, we have a meromorphic continuation of $\chi \td P_0(\lmb)^{-1} \chi: L^2(\calM) \to L^2(\calM)$ for $\lmb$ in \[
        \bigcup_{\bt \in (-\tfrac{\pi}{2}, \tfrac{\pi}{2})} \Lmb_\bt = \left\{\lmb \in \bbC \setminus \set{0} : \arg \lmb \in (-\frac{\pi}{2}, \pi + \frac{\pi}{2})\right\}.
    \]
    Since $\td P_0(\lmb)$ only differs from $P_0(\lmb)$ by multiplying a smooth function in a region away from the horizons, the meromorphic continuation applies to $\chi R(\lmb)\chi$.

    In what follows, we need to show that $\QNM(\bfr_1, M_1, \ldots, \bfr_N, M_N)$ is independent of the choice of $\chi \not\equiv 0$. It suffices to show that for every such $\chi \in C^\infty_{\rm{c}}$, we have \begin{equation}\label{eq:pole-kernel-relation}
        \lmb_0 \hbox{ is a pole of } \chi \td P_\bt(\lmb)^{-1} \chi \iff \ker \td P_\bt(\lmb_0) \neq \set{0} \hbox{ for some } \bt, \lmb_0 \in \Lmb_\bt. 
    \end{equation}
    The implication $\Rightarrow$ is obvious and we now prove the converse. 

    Suppose that $\ker \td P_\bt(\lmb_0) \neq \set{0} \hbox{ for some } \bt, \lmb_0 \in \Lmb_\bt$. Thanks to Lemma~\ref{lem:Pbt-inv}, \[
        \lmb \mapsto \td P_\bt(\lmb)^{-1}U^{4}: H^{0, (r_\infty, r_1, r_2, \ldots, r_N)}_\scop(\br{\calM_\bt}) \to H^{2, (r_\infty, r_1, r_2, \ldots, r_N)}_\scop(\br{\calM_\bt})
    \]
    is a meromorphic family of operators with poles of finite rank, i.e., there exist holomorphic operators $B_0(\lmb)$ and finite rank operators $B_k$ around $\lmb_0$ ($k = 1, \ldots, K(\lmb_0)$) such that  \[
        \td P_\bt(\lmb)^{-1}U^{4}(F_\bt(x)) = B_0(\lmb) + \sum_{k = 1}^{K}\frac{B_k}{(\lmb^2 - \lmb_0^2)^k} 
    \]
    holds near $\lmb_0$.
    Examining the procedure of complex scaling, we note that $\td P_\bt(\lmb)^{-1}U^{4}$ can be expressed as \[
        \td P_\bt(\lmb)^{-1}U^{4} = (\hat P_\bt - \lmb^2)^{-1},
    \]
    where $\hat P_\bt: H^{2, (r_\infty, r_1, r_2, \ldots, r_N)}_\scop(\br{\calM_\bt}) \to H^{0, (r_\infty, r_1, r_2, \ldots, r_N)}_\scop(\br{\calM_\bt})$ is an operator independent of $\lmb$. Therefore, it is standard that (see for instance \cite[Theorem 3.9]{DyZw19}) \[
        B_{k} = -(\hat P_\bt - \lmb_0^2)^{k-1}\Pi_{\lmb_0} = \frac1{2\pi i} \oint_{\lmb_0} (\hat P_\bt - \lmb^2)^{-1}2(\lambda^2-\lambda_0^2)^{k-1}\lmb\, \ud \lmb,
    \]
    where $B_{K + 1} = 0$. In particular, $(\hat P_\bt - \lmb_0^2) B_{K} = 0$. Let $m_{\lmb_0} := \mathrm{rank} \Pi_{\lmb_0} = \dim \mathrm{Ran} (\Pi_{\lmb_0}) < \infty$. Then $K(\lmb_0) \leq m_{\lmb_0}$.

    Suppose by contradiction that $\chi (\hat P_\bt - \lmb^2)^{-1} \chi$ is holomorphic near $\lmb_0$, then $\chi B_k \chi = 0$ for every $k \geq 1$. We claim that there exists $f$ such that $u = B_{K} \chi f$ is nontrivial. We first proceed to derive a contradiction from here and then prove this claim.
    
    Note that such a choice ensures \[
        \chi u = 0, \qquad (\hat P_\bt - \lmb_0^2) u = 0.
    \]
    Viewing $\hat P_\bt - \lmb_0^2$ as a second order elliptic operator on $\calM$ (see \eqref{eq:tdP-bt-lmb}), the principal part is $-U^{-4}\lap$ in $\set{r < R_{\infty, 1}} \cap \set{r_\bop > R_{\bop, 1}}$. Note that this region contains a point $x_0$ with $ \chi (x_0)\neq 0$ thanks to the choice in Lemma~\ref{lem:cutoff-res-bt-0}. In particular, $\hat P_\bt - \lmb_0^2$ has real-valued coefficient in its principal part at at least a point where $u$ vanishes nearby. Treating this as the base point, we apply the unique continuation theorem \cite[Theorem 17.2.6]{HormanderIII} and conclude that $u \equiv 0$ on $\calM$. Indeed, it is easy to check that the hypotheses in \cite[Theorem 17.2.6]{HormanderIII} are satisfied: the lower order terms of $\hat P_\bt - \lmb_0^2$ satisfy the estimates \cite[(17.2.2)']{HormanderIII} and the condition \cite[(17.2.9)]{HormanderIII} is trivial due to the vanishing of $u$ nearby. 
    This is a contradiction.
    
    Now it reduces to verify the claim above. In fact, it suffices to consider the dual problem and apply the unique continuation theorem to $(\hat P_\bt - \lmb_0^2)^*$ as above, where we recall that the rank of the dual is the same \cite[Chapter III \S 4.3]{Kato76}. This finishes the proof.
\end{proof}
\begin{remark}
    The contradiction argument above is akin to \cite[Proof of Theorem 1]{WunschZworski00} and \cite[Proof of Lemma 3.4]{Xiong22}. See also \cite{Sjostrand2000} for a treatment in the case of black box scattering.
\end{remark}

\begin{remark}\label{rmk:tht0-simpler-proof}
    Thanks to the conjugation \eqref{eq:conjugation-h}, meromorphic extension of $P(\lambda)^{-1}$ to a conic neighborhood of the real axis can be derived as a corollary of \cite{WunschZworski00}, where the size of the conic neighborhood is in turn determined by Lemma~\ref{lem:calU-set}.
\end{remark}

We record the intermediate step \eqref{eq:pole-kernel-relation} as a corollary, which gives an alternative definition of QNM.
\begin{corollary}[Quasinormal modes]\label{cor:QNM}
    $\QNM(\bfr_1, M_1, \ldots, \bfr_N, M_N)$ can be also characterized as the discrete subset of $\bbC \setminus i(-\infty, 0]$, which constitutes $\lmb \in \bbC \setminus i(-\infty, 0]$ such that 
    \[
         \hbox{ for some } \bt \in (-\tfrac{\pi}{2}, \tfrac{\pi}{2}) \hbox{ such that } \lmb \in \Lmb_\bt, \hbox{ we have } \ker \td P_\bt(\lmb) \neq \set{0}.
    \]
\end{corollary}

\begin{remark}\label{rem:lap}
Recall from \eqref{eq:conjugation-h} that $P(\lambda)= U^{3}(-\Delta_{h}-\lambda^2)U^{-1}$ where $h=U^4 \ud x^2$ is a scattering metric. By the limiting absorption principle for scattering metrics (see e.g., \cite[Proposition 14]{Melrose94} and \cite[Proposition 2.4]{Vasy98}), we have for $\lambda\in \mathbb{R}\setminus\{0\}$,
    \begin{equation*}
        \chi P(\lambda)^{-1}\chi = \lim\limits_{\eps\downarrow 0} \chi P(\lambda+i\eps)^{-1}\chi, \quad \chi \in C_{\rm{c}}^{\infty}(\calM),
    \end{equation*}
where the right-hand side converges in $L^2\to L^2$.
We conclude that there is no resonance on $\mathbb{R}\setminus\{0\}$.
\end{remark}

\section{Construction of quasimodes}\label{sec:quasimode}

\subsection{Structure of trapping}
We consider the operator \begin{equation}\label{eq:P_m_lmb}
    \begin{split}
        P_m(\lmb) &:= -\rho^{\frac12} e^{i\lmb t - i m \vtht} \Box_{g_{MP}} \rho^{-\frac12} e^{-i\lmb t + i m \vtht} \\
        &\ = \rho^{\frac12}e^{-i m \vtht} P(\lmb) \rho^{-\frac12}e^{i m \vtht} = -\rho^{\frac12}U^{-2}(\p_\rho^2 + \rho^{-1}\p_\rho - m^2 \rho^{-2} + \p_z^2) \rho^{-\frac12} - \lmb^2 U^2 \\
        &\ = -U^{-2}\lap_{\rho, z} + (m^2 - \frac14)\rho^{-2}U^{-2} - \lmb^2 U^2.
    \end{split}
\end{equation}
Note that $P_m(\lmb) u = 0$ is equivalent to \[
    -U^{-4} \lap_{\rho, z} u + (m^2 - \frac14) \rho^{-2} U^{-4}u = \lmb^2 u.
\]
Setting $h = (m^2 - \frac14)^{-\frac12}$, $\kpp = \lmb^2 h^2$, we have \begin{equation}\label{eq:e-val-formulation}
    (-U^{-4} h^2\lap_{\rho, z} + \rho^{-2} U^{-4}) u = \kpp u.
\end{equation}
To study such an eigenvalue problem, it is crucial to understand the structure of the potential $V = \rho^{-2} U^{-4}$. The importance of the function $h = \rho U^2$ can be seen from analyzing the geodesic flow as well \cite{ShipleyDolan16}. 
\begin{proposition}\label{prop:equal-mass-V-localmin}
    Assume that $M := M_1 = M_2$. The inequality $\frac{4}{\sqrt{27}} M < a < \frac{4\sqrt{2}}{\sqrt{27}} M$ holds if and only if the potential $V(\rho,z)$ admits a local minimum on $\mathbb{R}_{> 0}\times \mathbb{R}$. Moreover, whenever this condition is satisfied, the corresponding local minimum is unique.
\end{proposition}
\begin{proof}
    Set \begin{equation}\label{eq:h}
        h(\rho, z) := \rho U^2 = \rho \left( 1 + \frac{M}{\sqrt{\rho^2 + (z - a/2)^2}} + \frac{M}{\sqrt{\rho^2 + (z + a/2)^2}}\right)^2.
    \end{equation}
    We compute \[
        h_\rho = U(U + 2\rho U_\rho), \qquad h_z = 2\rho U U_z.
    \]
    Therefore, the critical points are characterized by \[\left\{\begin{split}
        &U = 2M\rho^2 \left( \frac{1}{\sqrt{\rho^2 + (z - a/2)^2}^3} + \frac{1}{\sqrt{\rho^2 + (z + a/2)^2}^3} \right), \\
        &\frac{z - a/2}{\sqrt{\rho^2 + (z - a/2)^2}^3} + \frac{z + a/2}{\sqrt{\rho^2 + (z + a/2)^2}^3} = 0.
    \end{split}\right.
    \]
    At the critical points, \begin{equation}\label{eq:Hess-comp}
        h_{\rho\rho} = U(3U_\rho + 2\rho U_{\rho\rho}) = \frac1{2\rho} U( 4\rho^2 U_{\rho\rho} - 3U),\qquad h_{\rho z} = 2\rho U U_{\rho z}, \qquad h_{zz} = 2\rho U U_{zz}.
    \end{equation}
    Introducing the renormalization \[
        \wh \rho := \frac{2\rho}{a}, \qquad \wh z := \frac{2z}{a}, \qquad \mu := \frac{M}{a}, \qquad U(\wh \rho, \wh z) = 1 + \frac{2\mu}{\sqrt{\wh \rho^2 + (\wh z - 1)^2}} + \frac{2\mu}{\sqrt{\wh \rho^2 + (\wh z + 1)^2}}, 
    \]
    the critical points are given by \begin{equation}\label{eq:crit-pt-renormalizedvar}
        \left\{\begin{split}
        &U + 2\wh \rho U_{\wh \rho} = 1 + \frac{2\mu}{\sqrt{\wh\rho^2 + (\wh z - 1)^2}} + \frac{2\mu}{\sqrt{\wh\rho^2 + (\wh z + 1)^2}} \\
        &\qquad\qquad\qquad\qquad\qquad\qquad\qquad\qquad- 4\mu\wh\rho^2 \left( \frac{1}{(\wh\rho^2 + (\wh z - 1)^2)^{\frac32}} + \frac{1}{(\wh\rho^2 + (\wh z + 1)^2)^{\frac32}} \right) = 0, \\
        &-\frac1{2\mu}U_{\wh z} = \frac{\wh z - 1}{(\wh\rho^2 + (\wh z - 1)^2)^{\frac32}} + \frac{\wh z + 1}{(\wh\rho^2 + (\wh z + 1)^2)^{\frac32}} = 0.
    \end{split}\right.
    \end{equation}
    {\it Case 1: Analyzing special points $(\wh\rho, \wh z) = (\wh\rho, 0)$. }
    It is easy to see that $\wh z = 0$ solves the second equation for all $\wh \rho$ and we examine such points first. Setting $\wh z = 0$ in the first equation, we have \begin{equation}\label{eq:mu-relation-crit}
        4\mu = \frac{(t+1)^{\frac32}}{t-1},
    \end{equation}
    where we set $t := \wh\rho^2$. Computing \begin{equation}\label{eq:mu-relation-crit-derivative}
        \frac{\ud}{\ud t}\left(\frac{(t+1)^{\frac32}}{t-1}\right) 
    = \frac12(t-1)^{-2}(t+1)^{\frac12}(t-5),
    \end{equation}
    we know that the existence of critical points of the form $(\wh\rho, \wh z = 0)$ requires the ratio condition $\mu \in [\frac{\sqrt{27}}{4\sqrt{2}}, \infty)$.

    In this case, one can read off from \eqref{eq:Hess-comp} that $h_{\rho z} = 0$. Therefore, it requires that $U_{zz} < 0$ and $3U_\rho + 2\rho U_{\rho\rho} < 0$. At $\wh z = 0$, $4\mu = \frac{(t+1)^{\frac32}}{t - 1}$, we have \[
    \left\{\begin{split}
        U_{\wh z \wh z} < 0 &\iff 2\left(\frac{1}{(\wh \rho^2 + 1)^{\frac32}} -  \frac{3}{(\wh\rho^2 + 1)^{\frac52}}\right) > 0 \iff t > 2, \\
        4\wh\rho^2 U_{\wh\rho\wh\rho} < 3U &\iff 4\wh\rho^2\left(\frac{-4\mu}{(\wh \rho^2 + 1)^{\frac32}} + \frac{12\mu \wh\rho^2}{(\wh \rho^2 + 1)^{\frac52}} \right) < 3 + \frac{12\mu}{\sqrt{\wh\rho^2 + 1}} \\
        &\iff 4t(8\mu t - 4\mu) < 3(t+1)^\frac{5}{2} + 12\mu(t+1)^2 \\
        &\iff 4t(2t - 1) < 3(t^2 - 1) + 3(t+1)^2
        \iff 2t^2 - 10t < 0 \iff t < 5.
    \end{split}\right.
    \]
    From \eqref{eq:mu-relation-crit}, we know that the existence of local maxima in this case is only possible when $\mu \in [\frac{\sqrt{27}}{4\sqrt{2}}, \frac{\sqrt{27}}{4}]$. Moreover, thanks to \eqref{eq:mu-relation-crit-derivative}, there exists a unique maximum if $\mu \in (\frac{\sqrt{27}}{4\sqrt{2}}, \frac{\sqrt{27}}{4})$ holds. We first proceed with justifying that no local maxima with $z \neq 0$ and then address the endpoint scenario.

    {\it Case 2: Any critical point with $z \neq 0$ cannot be a local maximum.}
    Without loss of generality, we may assume that $z>0$.
    Set \[
        q := \sqrt{\frac{\rho^2 + (z - a/2)^2}{\rho^2 + (z + a/2)^2}} \in (0, 1).
    \]
    $U_z = 0$ implies that \[
        \frac{1 - \wh z}{\wh z + 1} = q^3 \iff \wh z = \frac{1 - q^3}{1 + q^3}.
    \]
    Substituting this back to the definition of $q$, we have \begin{align*}
        q^2 - 1 = \frac{-4\wh z}{\wh \rho^2 + (\wh z + 1)^2} \iff \wh \rho^2 &= \frac{4(1 + q + q^2)}{(1+q^3)(1 + q)} - \frac{4}{(1 + q^3)^2} \\
        &= \frac{4(1 + q + q^2 + q^3 + q^4 + q^5) - 4 - 4q}{(1 + q^3)^2 (1 + q)} = 4\frac{q^2(1 + q^2)}{(1 + q^3)^2}.
    \end{align*}
    Therefore, the critical points lie on the curve parametrized by $q$: \[
        \wh\rho(q) = \frac{2q\sqrt{1 + q^2}}{1 + q^3}, \qquad \wh z(q) = \frac{1 - q^3}{1 + q^3}.
    \]
    Plugging this back into the first equation of \eqref{eq:crit-pt-renormalizedvar}, a direct calculation gives \begin{align*}
        \mu = \frac{q(1 + q^2 + q^4)^{\frac32}}{(1 + q^3)^2(1 - q + q^2)} =: \mu_*(q),
    \end{align*}
    where $q \neq 1$ is necessary in the reduction. 

    We claim that for each fixed $\mu$, for every $q \in (0, 1)$, \[
        \frac{\ud^2}{\ud q^2} h(\rho(q), z(q))|_{\mu= \mu_*(q)} > 0, \qquad \rho(q) := \frac{a}{2}\wh\rho(q),\quad z(q) := \frac{a}{2}\wh z(q).
    \]
    Therefore, we know that at each critical point, there is a direction such that $h$ is growing and thus cannot be a local maximum. Indeed, 
    we first observe from \eqref{eq:crit-pt-renormalizedvar} that the function $h_\rho(\rho(q), z(q))/U(\rho(q), z(q))$ is linear with respect to $\mu$ such that the constant term is $1$. Moreover, the function vanishes at $\mu = \mu_*(q)$. Thus, \[
        h_\rho(\rho(q), z(q)) = U(\rho(q), z(q))\left(1 - \frac{\mu}{\mu_*(q)}\right).
    \]
    and \[
        \frac{\ud}{\ud q} h(\rho(q), z(q)) = h_\rho(\rho(q), z(q)) \frac{\ud \rho}{\ud q} = U(\rho(q), z(q))(1 - \frac{\mu}{\mu_*(q)})\frac{\ud \rho}{\ud q}.
    \]
    Furthermore, \[
        \frac{\ud^2}{\ud q^2} h(\rho(q), z(q))|_{\mu = \mu_*(q)} = U(\rho(q), z(q))\mu_*(q)^{-1}\frac{\ud \mu_*(q)}{\ud q} \frac{\ud \rho}{\ud q} > 0,
    \]
    where the last step follows by a direct calculation verifying that the first order derivatives of both $\mu_*$ and $\wh\rho$ are positive in $q \in (0, 1)$. This finishes the proof. 

    {\it Case 3: Threshold cases $\mu = \frac{\sqrt{27}}{4\sqrt{2}}$ or $\frac{\sqrt{27}}{4}$. }
    The threshold value $\frac{\sqrt{27}}{4\sqrt{2}}$ corresponds to the degeneracy $t = 5$ in \eqref{eq:mu-relation-crit-derivative} and it is easy to see that it is not a local maximum. For the other threshold value, using the parameter $q$ introduced in Case~2, one can conclude that $h(\rho(q), z(q))$ is decreasing in $q$ similarly. Hence, the threshold values cannot be local maxima.
\end{proof}
\begin{remark}
    In fact, one could show that all such critical points in Case~2 are saddle points by verifying $U_{\wh z \wh z} < 0$. We omit the discussion here.
\end{remark}
\begin{remark}
    Thanks to \eqref{eq:mu-relation-crit}, there are two critical points on the equatorial plane ($z = 0$) when $\frac{a}{M} < \frac{4}{\sqrt{27}}$, i.e., $\mu > \frac{\sqrt{27}}{4}$. They correspond to the two roots $t_{1,\mu} < 5 < t_{2, \mu}$ of \eqref{eq:mu-relation-crit}. We expand for $t = t_{2,\mu} = \wh \rho_{2,\mu}^2$: \[
        16\mu^2 = \frac{(t+1)^3}{(t-1)^2} = (t-1) + 6 + \frac{12}{t-1} + \frac{8}{(t-1)^2}.
    \]
    Hence, as $\mu \to \infty$, we have $\frac{t_{2,\mu}}{16\mu^2} = \frac{\rho_{2,\mu}^2}{4M^2} \to 1$, i.e., $\rho_{2, \mu} \to 2M$, which matches the location of the photon sphere for the limiting extremal RN black hole with mass $2M$. We refer to \cite{OkOs24} for an account of this property by solving null geodesic equations numerically. 
\end{remark}

\subsection{Energy estimate and proof of Proposition~\ref{prop:quasimodes}}
We proceed under the assumptions of Proposition~\ref{prop:quasimodes}. In the preceding subsection, we reduced our problem to an eigenvalue problem with a potential well. Now we follow the framework introduced in \cite{HolzegelSmulevici14} and \cite{Benomio} in the discussion below. 
Before stating our main estimate, we need the following general fact.
\begin{lemma}
    Let $\Omg \Subset \R_{\rho> 0} \times \R_z$ be a bounded open set and $h \in \R_+$.
    Given smooth functions $u: \Omg \to \bbC$ such that $u|_{\p\Omg} = 0$ and functions $W \in C(\br\Omg; \R), \phi \in \mathrm{Lip}(\br\Omg; \R)$, we have \[
        \int_\Omg |h\nabla_{\rho, z} (e^{\frac{\phi}{h}} u)|^2 + (W - |\nabla_{\rho, z} \phi|^2) e^{2\frac{\phi}{h}} |u|^2\, \ud \rho \, \ud z = \Re\int_\Omg (-h^2\lap_{\rho, z} u + Wu) \br u e^{2\frac{\phi}{h}}\, \ud \rho\, \ud z,
    \]
    where $|\nabla_{\rho, z} v|^2 := |\p_\rho v|^2 + |\p_z v|^2$.
\end{lemma}
\begin{proof}
    It follows from a direct integration by parts and we omit the proof. See for instance \cite[Lemma 10.1]{Benomio}, \cite[Proposition 6.1]{TaoZworski} for a derivation.
\end{proof}

In what follows, we adopt the following setup.
In view of Proposition~\ref{prop:equal-mass-V-localmin}, we consider a fixed bounded open set $\Omg \Subset \R_{\rho > 0} \times \R_z$ containing the local minimum of $V = \rho^{-2} U^{-4}$. We fix a suitable energy level $E \in (\min_\Omg V, \min_{\p\Omg} V)$. Then for sufficiently small $\dlt$, for all $\kpp$ such that $|\kpp - E| < \dlt$, there exists $\dlt' \ll 1$ such that we have $V - \kpp \geq c$ for some $c > 0$ on \begin{equation}\label{eq:Omg_dlt_defn}
    \Omg_{\dlt'} := \set{p \in \Omg : \dist(p, \p\Omg) < \dlt'}, \qquad \dist(\cdot, \cdot)\hbox{ refers to the Euclidean coordinate distance}.
\end{equation}

Then for $\eps > 0$, we consider the following partition of $\Omg$: \[
    \Omg_\eps^+(E) := \set{p : U^4(p)(V(p) - E) \geq \eps} \cap \Omg, \qquad \Omg_\eps^-(E) := \set{p : U^4(p)(V(p) - E) < \eps} \cap \Omg.
\]

\begin{lemma}\label{lem:eng-est-u}
    Given an energy level $E \in (\min_\Omg V, \min_{\p\Omg} V)$. For any $\eps \ll 1$, there exists $\dlt = \dlt(\eps) \ll 1$ such that the following holds. For any $u$ solving the Dirichlet eigenvalue problem \eqref{eq:e-val-formulation} with eigenvalue $\kpp$ with $|\kpp - E| < \dlt$. We introduce $\alp_E(\eps) := \sup_{\Omg_\eps^-(E)} d_E$ and \[
        \phi_{E, \eps} := (1 - \eps) d_E.
    \] 
    Then we have \[
    \int_\Omg |h\nabla_{\rho, z} (e^{\frac{\phi_{E, \eps}}{h}} u)|^2\, \ud \rho\, \ud z + \frac12 \eps^2 \int_{\Omg_\eps^+(E)}  e^{2\frac{\phi_{E, \eps}}{h}} |u|^2\, \ud \rho \, \ud z \aleq_{\Omg, a, M} (\kpp + \frac12 \eps) e^{2\alp_E(\eps)/h} \|u\|_{L^2(\Omg)}^2. 
    \]
\end{lemma}
\begin{proof}
    We apply the previous lemma to $\phi(\rho, z) = \phi_{E, \eps}$ and $W := U^4(V - \kpp)$. Then we have \begin{equation}\label{eq:agmon-est-spe}
        \int_\Omg |h\nabla_{\rho, z} (e^{\frac{\phi}{h}} u)|^2 + \int_{\Omg_\eps^+} (W - |\nabla_{\rho, z} \phi|^2) e^{2\frac{\phi}{h}} |u|^2\, \ud \rho \, \ud z = \int_{\Omg_\eps^-}  (|\nabla_{\rho, z} \phi|^2 - W) e^{2\frac{\phi}{h}} |u|^2\, \ud \rho \, \ud z.
    \end{equation}
    On $\Omg_\eps^-$, we have \[
        -W = U^4(\kpp - V) \leq C \kpp,
    \]
    where $C$ depends on $\Omg, \eps, U$. On the other hand, thanks to Lemma~\ref{lem:Agmon-gradient}, it holds on $\Omg_\eps^-(E)$ that \[
        |\nabla_{\rho, z} \phi|^2 \leq (1-\eps)^2 |\nabla d_E|^2 \leq (1-\eps)^2 \eps \leq \eps (1- \eps).
    \]
    Thus, \[
        \hbox{RHS of \eqref{eq:agmon-est-spe}} \leq C(\kpp + \eps(1-\eps)) e^{2\frac{\alp_E(\eps)}{h}} \|u\|_{L^2(\Omg)}^2.
    \]
    On the other hand, we consider the integral on the left hand side over $\Omg_\eps^+$. On $\Omg_\eps^+$. By Lemma~\ref{lem:Agmon-gradient}, we have \[
        W - |\nabla_{\rho, z} \phi|^2 \geq U^4(V - E) - C \dlt - (1-\eps)^2|\nabla d_E|^2 \geq (2\eps - \eps^2) U^4(V - E) - C\dlt \geq \eps^2 - C \dlt,
    \]
    where $C$ is dependent on $\Omg, U$ and the last inequality requires $\eps \ll 1$. Now, choosing $\dlt \leq \eps^2/(2C)$, we finish the proof.
\end{proof}

Using the previous estimates, we are now ready to prove an estimate in the classically forbidden region. 

\begin{proposition}\label{prop:est-forb-region}
    Given an energy level $E \in (\min_\Omg V, \min_{\p\Omg} V)$. For $\dlt$ sufficiently small, there exists $\dlt' \ll 1$ such that there exists $m_0$ such that for all angular frequency $m \geq m_0$, there exists a solution $u_m$ of the Dirichlet eigenvalue problem \eqref{eq:e-val-formulation} with eigenvalue $\kpp_m = \lmb^2 (m^2 - \frac14)^{-1}$ with $|\kpp_m - E| < \dlt$ and $u_m$ satisfies the estimate \[
        \int_{\Omg_{\dlt'}} (|\nabla_{\rho, z} u_m|^2 + |u_m|^2) \, \ud \rho\, \ud z \leq Ce^{-Cm} \|u_m\|_{L^2(\Omg)}^2,
    \]
    where $\Omg_{\dlt'}$ is defined in \eqref{eq:Omg_dlt_defn} and $C$ is independent of $m$.
\end{proposition}
\begin{proof}
    By Weyl's law (Proposition~\ref{prop:weyl}), the existence of $u_m$ with eigenvalue in the window $(E - \dlt, E + \dlt)$ for $m$ large (i.e., $h$ small) follows.

    For sufficiently small $\dlt$, we choose $\eps \ll 1$ such that we can apply Lemma~\ref{lem:eng-est-u} to $u_m$ with $\dlt \leq \dlt(\eps)$. In addition, $\eps \ll 1$ is fixed such that $\dlt'$ is small enough such that there exists $c > 0$, $\phi_{E, \eps} \geq \frac12 d_E > c$ on $\Omg_{\dlt'}$, $\Omg_{\dlt'} \subseteq \Omg_{\eps}^+(E)$ and $\alp_E(\eps) = o(1) < \frac{c}{2}$. Since $\Omg_{\dlt'} \subseteq \Omg_{\eps}^+(E)$, we now have \[
         \int_{\Omg_{\dlt'}} |h\nabla_{\rho, z} (e^{\frac{\phi_{E, \eps}}{h}} u_m)|^2\, \ud \rho\, \ud z + \int_{\Omg_{\dlt'}} e^{2\frac{\phi_{E, \eps}}{h}} |u_m|^2\, \ud \rho \, \ud z \aleq e^{2\alp_E(\eps)/h} \|u_m\|_{L^2(\Omg)}^2, \qquad h^2 = (m^2 - \frac14)^{-1}.
    \]
    Expanding the gradient term using product rule, one derives by the uniform bound on $\nabla d_E$ that \[
        \int_{\Omg_{\dlt'}} |he^{\frac{\phi_{E, \eps}}{h}} \nabla_{\rho, z} u_m|^2\, \ud \rho\, \ud z + \int_{\Omg_{\dlt'}} e^{2\frac{\phi_{E, \eps}}{h}} |u_m|^2\, \ud \rho \, \ud z \aleq e^{2\alp_E(\eps)/h} \|u_m\|_{L^2(\Omg)}^2, \qquad h^2 = (m^2 - \frac14)^{-1}.
    \]
    In particular, this implies that \[
        \int_{\Omg_{\dlt'}} (|h\nabla_{\rho, z} u_m|^2 + |u_m|^2)\, \ud \rho \, \ud z \aleq e^{-c/h} \|u_m\|_{L^2(\Omg)}^2.
    \]
    Noting that $h^{-2} e^{-c/h} \leq e^{-c/2h}$ for $h \ll 1$, one concludes the proof.
\end{proof}

Now we can apply Proposition~\ref{prop:est-forb-region} to construct quasimodes. 
\begin{proof}[Proof of Proposition~\ref{prop:quasimodes}]
    We take the sequence $(u_m, \kpp_m, \lmb_m)$ constructed in Proposition~\ref{prop:est-forb-region}. We note that \[
        \frac{\lmb_m^2}{m^2} \sim \kpp_m \in (E - \dlt, E + \dlt),
    \] 
    the first property in the proposition holds true. 
    
    Let $\chi = \chi(\rho, z) \in C^\infty_{\rm{c}}(\Omg; \R)$ such that $\chi \equiv 1$ in $\Omg\setminus\Omg_{\dlt'}$. 
    We set \[
        v_m := \chi(\rho, z)\rho^{-\frac12} e^{im \vtht} u_m(\rho, z),
    \]
    which satisfies the second property as required. 
    Thanks to \eqref{eq:P_m_lmb}, one can derive from the explicit formula of $\mathrm{Err}(v_m) := P(\lmb_m) v_m$ that \[
        \|\mathrm{Err}(v_m)\|_{L^2(\calM)} \aleq \|u_m\|_{H^1(\Omg_{\dlt'})} \aleq e^{-Cm} \|v_m\|_{L^2(\calM)},
    \]
    where we apply Proposition~\ref{prop:est-forb-region} in the last step. Higher order estimates are also standard.
    By renormalizing $v_m$ in $L^2(\calM)$ and relabeling, we finish the proof.
\end{proof}


\section{From quasimodes to resonances: Proof of Theorem~\ref{thm:exp-conv-QNM}}

We first recall \cite[\S 4 Lemma 2]{TangZworski}; see also \cite{Stefanov99} for a restatement with remarks on the uniformity of constants.
\begin{lemma}\label{lem:TangZworski2}
    Let $h_j>0$ be a sequence with $h_j\to 0$. Let $F(z,h)$ be an operator-valued holomorphic function defined in a neighborhood of
    \begin{equation*}
        \Omega(h) := [E(h)-5h,E(h)+5h]+i [-S(h)h^{-\mathfrak{n}-1},S(h)]
    \end{equation*}
    for $h\in\{h_j\}$, $E(h)\in \mathbb{R}$ and $De^{-D/h}\leq  S(h)=O(h^{\infty})$.
    If $F(z,h)$ satisfies
    \begin{equation}\label{eq:norm-omega}
        \|F(z,h_j)\| \leq  A \exp \left(A h_j^{-\frkn}\log(1/(h_j S(h_j)))\right),\quad z\in \Omega(h_j).
    \end{equation}
    \begin{equation}\label{eq:norm->0}
        \|F(z,h_j)\|\leq \frac{A}{\Im z}, \quad  z\in \Omega(h_j), \, \Im z >0.
    \end{equation}
    Then there exists $h_0>0$ and $B>0$ (independent of $E(h)$ and $F$) such that
    \begin{equation*}
        \|F(z,h_j)\|\leq \frac{B}{S(h_j)},\quad z\in [E(h_j)-h_j,E(h_j)+h_j],\quad h_j\leq h_0.
    \end{equation*}
\end{lemma}

\begin{proof}[Proof of Theorem~\ref{thm:exp-conv-QNM}]
We prove by contradiction. Note that there is no resonance on $\bbR\setminus\{0\}$ by Remark~\ref{rem:lap}. Let $E>0$ be the same as in Proposition~\ref{prop:quasimodes} and $c>0$ be one half of the $c$ in Proposition~\ref{prop:quasimodes}. Take $\delta>0$ sufficiently small. Suppose there is no resonance in the region
\begin{equation*}
    [(\sqrt{E}-\delta)m , (\sqrt{E}+\delta)m]+ i[-e^{-cm},0].
\end{equation*}
Let $h_m=m^{-1}$. We consider the operator
\begin{equation*}
    F(z,h):=\chi (h^2P(h^{-1}z))^{-1}\chi =  \chi(h^2\td P_\bt(h^{-1}z))^{-1} U^2(x)\chi,
\end{equation*}
on the Hilbert space $L^2(\calM, U^4 \ud x)$, where $\bt \in (-\frac{\pi}{2}, \frac{\pi}{2})$ is to be determined. Then $F(z,h)$ is holomorphic on $[ \sqrt{E}-\delta , \sqrt{E}+\delta ]+i[-he^{-ch^{-1}},2]$ for $h \in \set{h_m}$. We note that the holomorphy in the upper half plane is automatic by \eqref{eq:U4-normalized-resolvent}. Let $S(h) :=\frac{1}{2} h^5 e^{-ch^{-1}}$. In particular, $F(z, h)$ is holomorphic on 
\begin{equation*}
    \Omega(h):=[\sqrt{E}-\delta , \sqrt{E}+\delta ]+i[-2S(h) h^{-4}, 2].
\end{equation*}
It is clear that \eqref{eq:norm->0} holds due to the self-adjointness of $-U^{-4}\Delta$. We claim that \eqref{eq:norm-omega} holds with $\mathfrak{n}=3$. Let $Q\in \Psi^{\mathrm{comp}}_h(\calM)$ be a compactly microlocalized finite rank self-adjoint operator such that $Q\geq 0$ and $\mathrm{rank}\, Q \lesssim h^{-3}$ (this can be constructed using functional calculus, see \cite[Theorem 14.9]{semibook}, \cite[Section 2.5]{JinZworski17}) and $Q$ is elliptic in a large enough compact set so that $h^2\td P_{\bt}(h^{-1}z) -iQ$ is semiclassically elliptic, where we fix $\bt$ such that $[\sqrt{E}-10\delta, \sqrt{E}+10\delta ]+i[-10\delta,10\delta] \subset \Lmb_\bt$. The ellipticity in the region $\{r<R_{\infty,2}\}\cap \{r_{\bop}>R_{\bop,2}\}$ is obvious by taking $Q\geq 0$ sufficiently large to cover those regions. In the regions $\{r\geq R_{\infty,2}\} $ or $ \{r_{\bop}\leq R_{\bop,2}\}$, $h^2\td P_{\bt}(h^{-1}z)$ is already semiclassically elliptic and the term $-iQ$ does not affect the ellipticity.

Then by the semiclassical elliptic estimate Proposition~\ref{prop:ell-semi}, we have
\begin{equation}\label{eq:semi-ell-1}
    \|(h^2\td P_{\bt}(h^{-1}z) - iQ)^{-1}\|_{L^2\to L^2 } \lesssim 1, \qquad z \in [\sqrt{E}-10\delta, \sqrt{E}+10\delta ]+i[-10\delta,10\delta].
\end{equation}
Since
\begin{equation}\label{eq:res-ident-1}
    h^2\td P_{\bt}(h^{-1}z) = (h^2\td P_{\bt}(h^{-1}z)-iQ)(I+i(h^2\td P_{\bt}(h^{-1}z)-iQ)^{-1}Q),
\end{equation}
it suffices to estimate $(I+G)^{-1}$ with $G:=i (h^2\td P_{\bt}(h^{-1}z)-iQ)^{-1}Q$. Since $\mathrm{rank}\,{G}\leq \mathrm{rank}\, Q \aleq h^{-3}$ and $\|G\|_{L^2\to L^2}\lesssim 1$, we have
\begin{equation*}
     |\det (I+G(z))| \leq (1 + \|G\|)^{\mathrm{rank}\, G} \leq \exp(C_1 h^{-3}), \quad z\in [\sqrt{E}-10\delta, \sqrt{E}+10\delta ]+i[-10\delta,10\delta].
\end{equation*}
Moreover, we note that $(I+G(z))^{-1}=I-i(h^2\td{P}_{\beta}(h^{-1}z))^{-1}Q$. By elliptic estimate Proposition~\ref{prop:ell-semi}, $\|(h^2\td{P}_{\beta}(h^{-1}z))^{-1}\| \aleq 1$ for $z\in [\sqrt{E}-10\delta, \sqrt{E}+10\delta ]+i[\delta,10\delta] $, and
\begin{equation}
    |\det (I+G(z))^{-1}|\leq e^{C_1h^{-3}},\quad z\in [\sqrt{E}-10\delta, \sqrt{E}+10\delta ]+i[\delta,10\delta] .
\end{equation}
By Jensen's formula, \begin{equation*}
    \begin{split}
        &\peq \# \hbox{ zeros of $\det(I+G(z))$ inside $[\sqrt{E}-2\delta, \sqrt{E}+2\delta ]+i[-2\delta,2\delta]$} \\
        &\aleq \sup_{ [\sqrt{E}-10\delta, \sqrt{E}+10\delta ]+i[-10\delta,10\delta]} \log \left| \det(I+G(\cdot))\right| + \sup_{[\sqrt{E}-10\delta, \sqrt{E}+10\delta ]+i[\delta,10\delta]} - \log \left|\det(I+G(\cdot))\right|\\
        &\leq C_2h^{-3}.
    \end{split}    
\end{equation*}

Suppose $\det(I+G(z))=\td{G}(z)\prod(z-z_\ell)$ where $z_{\ell}$ are its zeros in $[\sqrt{E}-2\delta, \sqrt{E}+2\delta ]+i[-2\delta,2\delta]$. By \cite[Lemma 8.1]{Sjostrand97} and the arguments after it, we have
\begin{equation*}
    |\td{G}(z)|\leq e^{C_3h^{-3}}, \quad z\in [\sqrt{E}-2\delta, \sqrt{E}+2\delta ]+i[-2\delta,2\delta].
\end{equation*}
By Harnack's inequality applied to $C_3h^{-3}-\log|\td{G}(z)|$, we know that
\begin{equation*}
    -\log|\td{G}(z)|\leq C_4 h^{-3},\quad z\in [\sqrt{E}-\delta, \sqrt{E}+\delta ]+i[-\delta,\delta].
\end{equation*} 
Let \[\td{\Omega}(h):=[\sqrt{E}-\delta/2, \sqrt{E}+\delta/2]+i[- h^{-4}S(h),S(h)] \subseteq [\sqrt{E}-\delta, \sqrt{E}+\delta ]+i[-\delta,\delta].\]
For any $z\in \td \Omg(h)$, by the holomorphy of $F$ on $\Omg(h)$ and Corollary~\ref{cor:QNM}, we know that the zeros of $\det(I+G(z))$ are located outside the $S(h)$ neighborhood of $z$, and hence 
\begin{equation}\label{eq:logdet-IG-lower}
\begin{split}
    \log|\det(I+G(z))| & \geq \log|\td{G}(z)| + \sum_{\ell} \log|z-z_{\ell}|
    \\
    &\geq -C_4 h^{-3}- C_2h^{-3}\log(1/S(h)),\quad z \in \td{\Omega}(h).
\end{split}
\end{equation}
By \eqref{eq:semi-ell-1} and \eqref{eq:res-ident-1}, this implies that 
\begin{equation*}
    \|F(z,h)\|_{L^2\to L^2}\lesssim \|(I+G)^{-1}\|_{L^2\to L^2}\lesssim \exp({C_5 h^{-3}}\log(1/S(h))),\quad z\in\td{\Omega}(h),
\end{equation*}
where the last inequality follows from the adjugate matrix formula and \eqref{eq:logdet-IG-lower}.

By partitioning $\td{\Omg}(h)$ into $O(h^{-1})$ subregions and applying Lemma~\ref{lem:TangZworski2} to each one, we obtain
\begin{equation*}
    \|F(z,h_m)\| \leq C_6m^5 e^{c m}, \quad z \in  [\sqrt{E}-\delta/4, \sqrt{E}+\delta/4] \subseteq \R.
\end{equation*}
This contradicts the quasimodes constructed in Proposition~\ref{prop:quasimodes}.
\end{proof}

\bibliographystyle{alpha}
\bibliography{bibliography}

\end{document}